# The nature is simple in essence

## - on gravitation and electromagnetism unification -

Viorel Drafta

drafta_v@yahoo.com

**Abstract**

The proposed temporal fluctuations model attempt for a unitary vision on gravity, electromagnetism and inertia. On obtain Newton law of gravitation and Coulomb law by starting from simple principles. On obtain too the main results of General Relativity Theory and Electromagnetism. One of the major points of present work is to obtain unified potential, which allow a unitary description of gravity and electromagnetism. As consequence, on obtain supplemental effects responsible for conservation laws violation in certain closed systems. On maintain the validity of conservation laws only if those systems are coupling with vacuum. This fact can be used for energy production and/or transport. This proposed model opening to quantum and nuclear physics, but those subjects will be developed in a future work.

### 1. Introduction

Due to similarity of laws for *long-range* interactions gravitation and electrostatics, physicists searched long time their common basis. Introducing space-time curvature, GRT succeed to describe gravitational interactions and resulting movement in an elegant way: *Matter tells space how to curve, space tells matter how to move* [Wheeler]. The attempt to introduce electromagnetism in tensorial formulation frame of GRT has not succeeded, although introduction of supplementary dimensions for space-time in Kaluza-Klein type theories was a promising way. It misses too - at least to present day - the attempt to develop a quantum gravity theory, following the model of quantum electrodynamics.

An alternate attempt to GRT is the theory of vacuum polarization - PV [35]. Introduced by Wilson and developed by Dicke, PV treats the metric changes by equivalent changes of electric permittivity and magnetic permeability of vacuum. Following, the Maxwell's equations in curved spaces are linked with refraction index change in vicinity of massive bodies.

Unruh [42] argue that metric changes in vicinity of massive bodies are due to proportionality of event duration with gravitational potential. This fact is experimentally confirmed.

This paper broach gravity and electromagnetism unification by starting from [35] and [42], being a revised and improved version of paper [11]. On obtain Newton law of gravitation, Coulomb law, the main results of GRT and Electromagnetism starting from simple principles. As consequence, on obtain supplemental effects responsible for conservation laws violation in certain closed systems. This opened the possibility of future technological developments.

### 2. Temporal Fluctuations Model

For a unitary vision of reality, it is natural to suppose that elementary constituents of matter must have the simplest and fewest properties. They must interact in the simplest way, too. As result, the whole world is a natural consequence of properties and interactions of elementary constituents. As Yukawa wrote: *Nature is simple in essence*. The present model is conceived to correspond as much as it is possible to this paradigm.

## 2.1 Temporal Fluctuations. Properties

Assuming the SRT postulates, the basis for proposed model is the hypothesis of elementary constituents existence (named temporal fluctuations - TF), with following properties:

1) space-time widespread. The TF dimension and duration must be smaller than for elementary particles, probably in range of Planck length and duration;

2) have the light speed c in every reference frame;

3) during their movements TF penetrate each other. Lifetime of one fluctuation depends on concentration of superposed fluctuations that have the same velocity orientation. *Birth* and *decay* rate of one fluctuation depends too on concentration of superposed fluctuations that have the same velocity orientation;

4) on suppose the linear superposition of reciprocal effects of fluctuations. That's why on consider that probability of fluctuation birth in a given volume and with a given orientation, is proportional with the concentration of fluctuations that have the same orientation. We'll see that linear fields superposition derive from this hypothesis, although it can be incorrect for large concentrations of fluctuations and/or strongly anisotropy distribution.

Shortly, on call temporal fluctuations those entities that penetrate each other and that influence each other their lifetime and their rate of birth.

On nominate fluctuation's concentration by n. The law for distribution of fluctuations by orientation must be likely:

$$dn(\phi, \theta) = C \cdot f(\phi, \theta) d\phi \, d\theta \tag{1}$$

where $f(\phi,\theta)$ is distribution's function of fluctuations with $(\phi,\theta)$ orientation and C is a constant. The value of C on obtain by normalization:

$$n = \int_{\phi, \theta} C \cdot f(\phi, \theta) d\phi \, d\theta \tag{2}$$

$\phi \in [0, 2\pi]$, and $\theta \in [0, \pi]$. It result:

$$dn(\phi, \theta) = \frac{n}{\int_{\phi, \theta} f(\phi, \theta) d\phi \, d\theta} \cdot f(\phi, \theta) d\phi \, d\theta \tag{3}$$

On consider that in vacuum far away from any particle or field the fluctuations have an isotropy distribution by velocity orientation and concentration of fluctuations is $n_0$. In this situation $f(\phi,\theta)=1$ and distribution law could be write:

$$dn_0(\phi, \theta) = \frac{n_0}{2\pi^2} d\phi \, d\theta \tag{4}$$

Because the rate of *birth* and *decay* are link with fluctuation's average lifetime, in a given volume the concentration of fluctuations is stationary. Increasing of concentration determines that fluctuation's average lifetime increase and, in the same time, the rate of *birth* and *decay* decreases so the new value of concentration remains the same. The above statement is correct for anisotropy fluctuation's distribution too. As result on considers that the whole system of fluctuations tend to self-preservation.

In third and fourth of above properties on suppose that fluctuations interact one with other only if their velocities have the same orientation. On argue this by starting to write the relative velocity (relativistic case) between two particles:

$$v_1'^2 = \frac{(\vec{v}_1 - \vec{v}_2)^2 - \frac{1}{c^2}(\vec{v}_1 \times \vec{v}_2)^2}{\left(1 - \vec{v}_1 \cdot \vec{v}_2 / c^2\right)^2} \tag{5}$$

where $v_1$ and $v_2$ are the velocities of the two particles in laboratory reference frame and $v_1'$ is the velocity of first particle relative to the second one.

If $v_1=v_2=c$, then for any nonzero angle between particles results $v_1'=c$. If the angle equals zero then $v_1'$ is indeterminate. If angle equals zero, on take $v_1=v_2=c'$ ( $c'<c$ ). In this situation, for every $c'<c$ results $v_1'=0$. If $c'$ strive to c, at limit it must $v_1'=0$ for $c'=c$.

In conclusion, for fluctuations with same orientation the relative velocities equals 0 and for the others situations is c. Considering that the interactions between fluctuations depends on superposition duration, it is natural to consider that for fluctuations with same orientation the interaction is strongly than in others situations. At limit, on consider that only the fluctuations with same orientation interact.

On tell in advance that elementary particles result by fluctuations self-organization. Because fluctuation's average lifetime is much more less than for the particle, the structure of the last one maintain although fluctuations that compose it at a given moment disappear and are replaced with another. The particle modifies both concentration and isotropy of distribution for fluctuations in the surrounding environment. Those modifications propagate with c velocity. We'll see that distant effects of those modifications determine the gravitational and electromagnetic interactions.

Because particles are composed from fluctuations and because SRT postulates are valid, on consider that fluctuations from particle's structure interact in the same manner with the leftover fluctuations, whatever is particle's velocity relative to fluctuations. This fact implies that it is no absolute reference frame exists.

It must emphasize that event duration for particles interactions is linked with average lifetime of fluctuations.

**2.2. Relationship between concentration and distribution's asymmetry of fluctuations with distance**

Let it be a region with isotropy distribution of fluctuations that have concentration $n_0$. Be a volume *V* in that region. The get out flux of fluctuations equals the get in flux of fluctuations.

If inside *V* the concentration of fluctuations is $n>n_0$ then the get out flux of fluctuations is bigger than the get in flux of fluctuations.

On the other hand, because of different concentration of fluctuations in the two regions, fluctuation's lifetime is greater inside *V* then outside. By passing from one region to another, the fluctuations suffer a process similar to refraction and *refractive index* is proportional to inverse of fluctuation's concentration. Consequently, a system of fluctuations (such a particle) suffers a drift motion towards the zone with greater lifetime. The *duration gradient* can be produced by a concentration gradient or/and anisotropy gradient.

Let it be a source particle in vacuum with isotropy concentration distribution of fluctuations $n_0$ at particle's surface. As distance R from particle increases, the concentration of fluctuations decreases so at infinite it becomes $n_0$. For find the dependence of concentration by distance, it is natural to suppose that flow of fluctuations is given by $\Phi=C_1 n$. On the other hand, the flow is the number of fluctuations passing in time unit through surface S:

$$\Phi = -\frac{dn}{dt} \cdot S \tag{6}$$

On suppose that particle's shape is spherical and as result the surfaces of equal concentration are spherical layers. Because fluctuations propagate with speed c on can write: $dt=dR/c$. Results:

$$\frac{dn}{n} = -\frac{C_1 \cdot c}{4p} \frac{dR}{R^2} \tag{7}$$

by imposing condition that $n \to n_0$ for $R \to \infty$ it results:

$$n = n_0 e^{\frac{q}{R}} \tag{8}$$

where $q=C_1 c/4\pi$.

Supposing that at surface of all type of particles the concentration of fluctuations is the same, the only difference between particles consists in their radius. On will see that this supposition lead to a link between mass and radius of particles, to understand the nature of inertia and to the Newton's gravitational attraction law.

In this situation on write $q=CR_s$ where C is constant and $R_s$ is radius of source particle. Relationship (8) can be writing:

$$n = n_0 e^{C\frac{R_s}{R}} \tag{9}$$

On any particle's surface ($R=R_s$) the concentration of fluctuations is:

$$n_s = n_0 e^C \tag{10}$$

At the particle's surface it is possible that fluctuations with concentration $n_s$ have no isotropy distribution of velocity orientation. If maximum anisotropy is on radial direction, there are two possibilities: the anisotropy maximum is oriented towards inside or outside of particle which correspond to positive and negative electrical charges. We'll see indeed that by introducing the asymmetry of fluctuation's distribution on obtain a kind of interaction between particles that have the behavior of electrostatic interaction.

On define the asymmetry ($\delta$) as ratio between concentration of fluctuations on maximum asymmetry direction and whole fluctuations concentration: $\delta = \Delta n/n$, where $\Delta n$ is from relationship (3). At surface of any charged particle the asymmetry is maximum and on write down $\Delta$.

The asymmetry from one region must propagate towards neighborhood regions. Due to property of self-preservation for any fluctuation's system, it is natural to consider that asymmetry flow is constant. On write:

$$\Phi_{\text{particle's surface}} = \Phi_{\text{at distance R}} \tag{11}$$

on radial direction it results:

$$\Delta \cdot \frac{const}{\Delta t} = d \cdot \frac{4\pi R^2}{\Delta t} \tag{12}$$

and finally:

$$d = \Delta \cdot \frac{ct}{R^2} \tag{13}$$

where $ct$ is constant having dimension of square of distance. In a point at distance $ct^{1/2}$ above particle's surface the asymmetry must be maximum: $\delta = \Delta$. Results that quantity $R_0 = ct^{1/2}$ signify a minimum distance for that kind of interaction.

For the fluctuations with velocity oriented towards particle on explain the propagation of interaction by considering that in fluctuation's lifetime the average walk is smaller than fluctuation's dimension.

### 2.3. Distribution of fluctuations by orientation

In figure 1 on take Ox as maximum asymmetry axis:

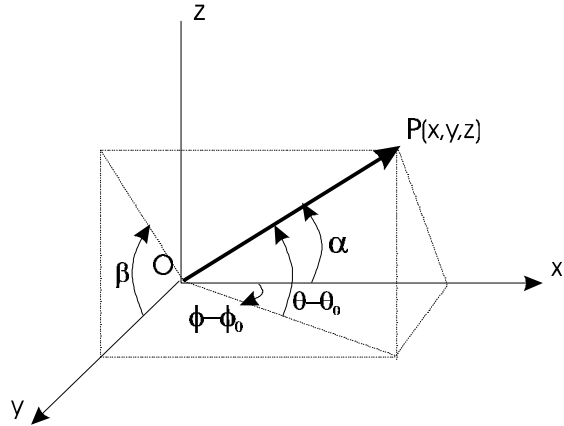

Figure 1

The pair of angles $\{\alpha \in [0,\pi], \beta \in [0,2\pi]\}$ and $\{\theta - \theta_0 \in [0,\pi], \phi - \phi_0 \in [0,2\pi]\}$ characterizes the orientation. If direction of maximum asymmetry does not coincide with Ox, then $\theta_0$ and $\phi_0$ are nonzero. If direction of maximum asymmetry coincide with Ox, then $\theta_0 = \phi_0 = 0$. On see:

$$\cos\alpha = \cos(\phi - \phi_0) \cdot \cos(\theta - \theta_0) \tag{14}$$

In #2.1 was introduced distribution's function by orientation. This function must depend on asymmetry $\delta$ and on angles. By using the pair of angles $\{\alpha, \beta\}$, the function must have the following behavior:

- for δ=Δ and α=0     ⇒ f(α)maximum;
- for δ=Δ and α=π     ⇒ f(α)=0;
- at great distance (δ=0)     ⇒ f(α)=1;
- function is continue in δ and α.

The function's graphic could look alike:

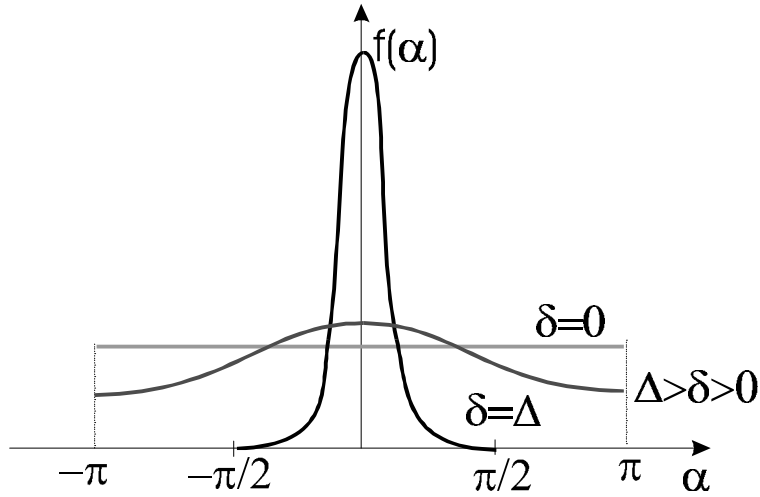

Figure 2

On see that for δ=Δ the asymmetry is maximum and fluctuations are oriented only towards outside or inside particle, depend on their charge.

A function that approximate the above graphic especial for δ<<Δ is:

$$f(\mathbf{a}) = e^{b \cdot \mathbf{d} \cdot \cos(\mathbf{a})} \tag{15}$$

where b is constant. Using pair of angles {ϕ, θ} on express the distribution function as:

$$f(\mathbf{j}, \mathbf{q}) = e^{b \cdot \mathbf{d} \cdot \cos(\mathbf{j} - \mathbf{j}_0) \cdot \cos(\mathbf{q} - \mathbf{q}_0)} \tag{16}$$

The relationship (16) allows choosing the most convenient axis system.
On see that by means of (15) and (16) we'll find later the Coulomb law.

**2.4. Interaction of fluctuations. Interaction potential**

Let it be a very small volume with concentration of fluctuations n (relationship (9)) and the asymmetry of distribution given by f (relationship (16)). Let it be the fluctuation j with orientation [ϕ, θ] relative to maximum asymmetry orientation [ϕ$_0$, θ$_0$]. Fluctuation j will interact with others Δn(ϕ, θ) fluctuations that have the same orientation. Quantity Δn(ϕ, θ) is given by (resulting from (3)):

$$\Delta n(\mathbf{j}, \mathbf{q}) = \frac{n}{\int_{\mathbf{j}, \mathbf{q}} f(\mathbf{j}, \mathbf{q}) d\mathbf{j}\, d\mathbf{q}} \cdot f(\mathbf{j}, \mathbf{q}) \Delta \mathbf{j}\, \Delta \mathbf{q} \tag{17}$$

On observe that for a finite value for Δn(ϕ, θ) it is necessary that Δϕ and Δθ do not have values that strive to 0. This is the reason for quantification of orientation. Another reason for quantification of orientation is the fact that only fluctuations with the same orientation interact.

The effect of Δn(ϕ, θ) fluctuations on j fluctuation is given by:

$$I_j(\mathbf{j}, \mathbf{q}) = a \cdot \mathbf{t} \cdot \Delta n(\mathbf{j}, \mathbf{q}) \tag{18}$$

where a is constant link with strength of interaction between fluctuations, and τ is the average lifetime of fluctuations in vacuum.

The average of effects on all fluctuations from the small region at the same time is:

$$I = \frac{1}{n} \int_{j,q} I_j(\boldsymbol{j},\boldsymbol{q}) dn \tag{19}$$

where dn is given by (3). On nominate quantity I the interaction potential. From I we'll derive gravitational and electric potentials. On find for I:

$$I = \frac{a \cdot \boldsymbol{t} \cdot n \cdot \Delta \boldsymbol{j} \cdot \Delta \boldsymbol{q}}{\left(\int_{j,q} f(\boldsymbol{j},\boldsymbol{q}) d\boldsymbol{j} d\boldsymbol{q}\right)^2} \cdot \int_{j,q} f^2(\boldsymbol{j},\boldsymbol{q}) d\boldsymbol{j} d\boldsymbol{q} \tag{20}$$

On replace f in above relationship with (16). By integrate between limits $\phi \in [0,2\pi]$ and $\theta - \theta_0 \in [0,\pi]$, on solve the integral by series development. The terms with odd powers of cosine are zero, so on obtain:

$$\int_{j,q} f(\boldsymbol{j},\boldsymbol{q}) d\boldsymbol{j} d\boldsymbol{q} = 2p^2 \left[ 1 + \left(\frac{b \cdot \boldsymbol{d}}{2\sqrt{2}}\right)^2 + \frac{1}{2} \cdot \frac{3}{4} \cdot \left(\frac{b \cdot \boldsymbol{d}}{2\sqrt{2}}\right)^4 + \ldots \right] \cong 2p^2 e^{\left(\frac{b \cdot \boldsymbol{d}}{2\sqrt{2}}\right)^2} \tag{21}$$

On see that the right sides approximate very well the exponent function. However, in weak field approximation (δ very small) the first order approximation in (21) will be enough for most of calculations.

In the same way:

$$\int_{j,q} f^2(\boldsymbol{j},\boldsymbol{q}) d\boldsymbol{j} d\boldsymbol{q} = 2p^2 \left[ 1 + \left(\frac{b \cdot \boldsymbol{d}}{\sqrt{2}}\right)^2 + \frac{1}{2} \cdot \frac{3}{4} \cdot \left(\frac{b \cdot \boldsymbol{d}}{\sqrt{2}}\right)^4 + \ldots \right] \cong 2p^2 e^{\left(\frac{b \cdot \boldsymbol{d}}{\sqrt{2}}\right)^2} \tag{22}$$

It results:

$$I = \frac{a \cdot \boldsymbol{t} \cdot n \cdot \Delta \boldsymbol{j} \cdot \Delta \boldsymbol{q}}{2p^2} \cdot e^{\frac{b^2 d^2}{4}} \tag{23}$$

Using (9) and (13) with notation:

$A = n_0 \, a \, \tau \, \Delta\phi \, \Delta\theta / 2\pi^2$

and

$B = b^2/4$ \hfill (24)

It result for interaction potential:

$$I = A \cdot e^{C \frac{R_s}{R} + B\Delta^2 \frac{R_0^4}{R^4}} \tag{25}$$

At very great distance from source the interaction potential is:

$I_0 = A$ \hfill (26)

### 2.5. Superposition

### 2.5.1 Superposition for concentration of fluctuations

Let it be two neutral particles with radius $R_{s1}$ and $R_{s2}$ that correspond to mass $m_1$ and $m_2$ (as we'll see later). For very closed particles the interaction potential of the two particles in P at great distance must equals the potential of a particle with mass $m = m_1 + m_2$ and radius $R = R_{s1} + R_{s2}$. In approximation of first degree, the concentration of fluctuations in P is the amount of $n_0$ with concentration excess that every particle produces:

$$n = n_0 \left( 1 + C \frac{R_{s1}}{R} + C \frac{R_{s2}}{R} \right) \cong n_0 e^{C \frac{R_{s1}}{R} + C \frac{R_{s2}}{R}} \tag{27}$$

As result, the interaction potential is:

$$I = A \cdot e^{C \frac{R_{s1}+R_{s2}}{R}} \tag{28}$$

It must emphasize that only in weak field approximation on have linear superposition, i.e. for $CR_s/R \ll 1$. Relationship (28) can be generalized for multiple sources:

$$I = A \cdot e^{C \sum_i \frac{R_{si}}{R_i}} \tag{29}$$

**2.5.2. Superposition of asymmetries**

Let it be two sources, which creates asymmetry of fluctuation's distribution in P. On choose a reference frame that maximum asymmetry directions for the two sources are $(\phi_1, \theta_0)$ and $(\phi_2, \theta_0)$. The two distribution functions are:

$$\begin{aligned} f_1 &= e^{b d_1 \cos(\varphi - \varphi_1) \cdot \cos(\theta - \theta_0)} \\ f_2 &= e^{b d_2 \cos(\varphi - \varphi_2) \cdot \cos(\theta - \theta_0)} \end{aligned} \tag{30}$$

The asymmetries are superpose in P by any direction $(\phi,\theta)$ so on write:

$$f = f_1 \cdot f_2 = e^{b \cdot \cos(\theta - \theta_0) \cdot [d_1 \cos(\varphi - \varphi_1) + d_2 \cos(\varphi - \varphi_2)]} \tag{31}$$

Following the demonstration from #2.4 with function of distribution given by (31), instead relationship (23) results:

$$I = \frac{a \cdot t \cdot n \cdot \Delta \varphi \cdot \Delta \theta}{2\pi^2} \cdot e^{\frac{b^2 \cdot [d_1^2 + d_2^2 + 2d_1 d_2 \cos(\varphi_1 - \varphi_2)]}{4}} = \frac{a \cdot t \cdot n \cdot \Delta \varphi \cdot \Delta \theta}{2\pi^2} \cdot e^{\frac{b^2 \cdot [\vec{d_1} + \vec{d_2}]^2}{4}} \tag{32}$$

On see that linearity of superposition take place only for very small $\delta$ (great distance from source) so the first order approximation it is valid.
Relationship (32) can be generalized for multiple sources:

$$I = \frac{a \cdot t \cdot n \cdot \Delta \varphi \cdot \Delta \theta}{2\pi^2} \cdot e^{\frac{b^2 \cdot \left[\sum_i \vec{d_i}\right]^2}{4}} \tag{33}$$

**2.6. Asymmetry due to a spherical source**

Let it be a spherical charged particle (electron or proton) and a point P at great distance from charge. The asymmetry in P is due to superposition of asymmetries produces by every point from particle's surface (see fig. 3).

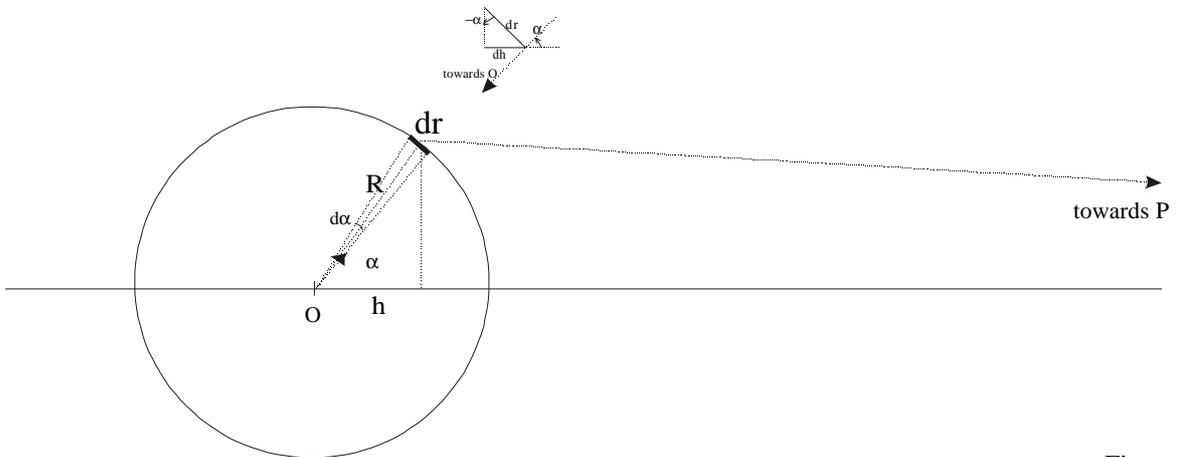

Figure 3

If $R_{OP}$ is large enough, then dr produces asymmetry $\delta\cos\alpha$ in P. The spherical crown with dr width produce in P the asymmetry $2\pi\delta\cos\alpha$. In the same time $dh/dr = -\sin\alpha$ and because $dr = Rd\alpha$ results $d\alpha = -dh/R\sin\alpha$. The resulting asymmetry is:

$$d_{rez} = 2pd \int_0^{\pi/2} \cos\alpha \, d\alpha = -\frac{2pd}{R}\int_0^R \frac{\cos\alpha}{\sin\alpha} dh = -\frac{2pd}{R}\int_0^R \frac{h}{\sqrt{R^2 - h^2}} dh = 2pd \qquad (34)$$

It must emphasize that the above result is valid only at great distance from source particle thus all contributions $\delta\cos\alpha$ can be considered parallel for any $\alpha$.

With above result, the interaction potential (25) produced in P by a charged particle stand still in O can be write:

$$I = A \cdot e^{C\frac{R_s}{R_{OP}} + 4p^2 B\Delta^2 \left(\frac{R_0^2}{R_{OP}^2}\right)^2} \qquad \text{where} \quad d = \Delta \frac{R_0^2}{R_{OP}^2} \qquad (35)$$

### 3. Interaction between particles

We are now able to treat the interaction between two particles. From above on see that, at particles surface, the concentration of fluctuations is $n_0 e^C$ and that decrease with distance increase. If particle's shape is spherical, its radius characterizes every type of particle. For charged particles, the distribution of fluctuations orientation is asymmetrical with maximum on radial direction. The anisotropy maximum is oriented towards inside or outside of particle, which correspond to positive and negative electrical charges. In present stage of theory development, it is impossible to say what asymmetry type correlate with positive charge and what asymmetry type correlate with negative charge. On see that it is attraction between particles with opposite fluctuation orientation and repel between particles with the same fluctuation orientation. For the particles with same radius exists the two asymmetry orientations, which correspond for particle and antiparticle. We found too that asymmetry decrease with square of distance.

In our argumentation on deal with two stable particles type: source particle (with radius $R_s$) and probe particle (with radius $r_0$). At both particles surface the asymmetry is the same, with maximum value $\Delta$.

On take the two interacting particles on Ox axis and on make convention that asymmetry is positive if is directed towards outside and negative if is directed towards inside. Angles are measured in trigonometric sense.

Source particle produces the fluctuations gradient and asymmetry gradient – $\delta$, and both of them produces a gradient of interaction potential of source particle. Due to symmetry of particles configuration, the gradients are oriented along Ox. On show at #2.2 that both concentration gradient and asymmetry distribution gradient involves the correspondent gradient of average lifetime of fluctuations, and particles move in direction of increasing the lifetime. We'll see that the value of interaction potential gradient is exactly the acceleration of probe particle due to concentration and asymmetry gradient produces by source particle.

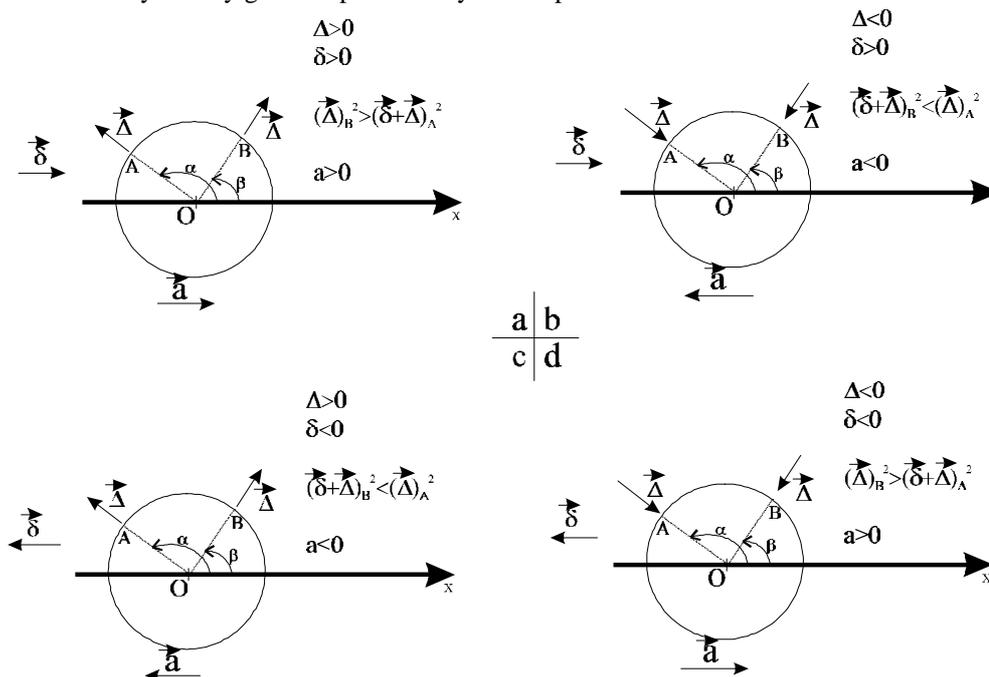

Figure 4

The concentration gradient always makes probe to move towards source particle and asymmetry that source produces superpose with asymmetry from probe particle surface. The value and orientation of probe acceleration depends on value and orientation of resulting gradient of asymmetry (see figure 4 for the four possible occurrence).

In O is probe particle and the source particle is somewhere at great distance on negative side of Ox axis. Figures 4a and 4d are for the interaction between particles with opposite charge while 4b and 4c are for the interaction between particles with same charge. It must emphasize that because asymmetry at particle's surface is maximum, the vectorial addition of $\delta$ and $\Delta$ take place only when absolute value of resultant is smaller than $\Delta$.

Let us now find the distance between two points located on a sphere with radius $r_0$.

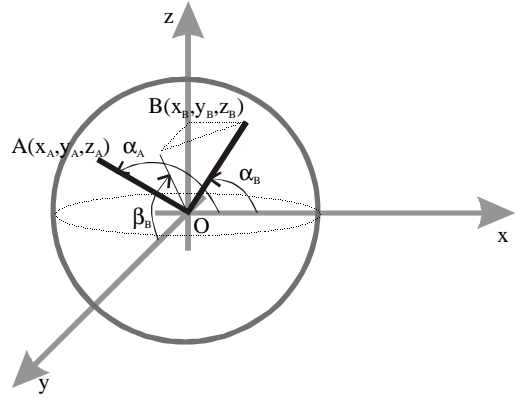

Figure 5

Using the notations from figure 5 ($\beta_A$ is not represented in figure) on write:

$$\overline{AB} = \left[(x_A - x_B)^2 + (y_A - y_B)^2 + (z_A - z_B)^2\right]^{\frac{1}{2}} \quad (36)$$

where:

$$\begin{aligned} x_{A,B} &= r_0 \cdot \cos\alpha_{A,B} \\ y_{A,B} &= r_0 \cdot \sin\alpha_{A,B} \cos\beta_{A,B} \\ z_{A,B} &= r_0 \cdot \sin\alpha_{A,B} \sin\beta_{A,B} \end{aligned} \quad (37)$$

In (36) on replace (37) and on obtain:

$$\overline{AB} = \sqrt{2} \cdot r_0 \cdot \sqrt{1 - \cos\alpha_A \cos\alpha_B - \sin\alpha_A \sin\alpha_B \cos(\beta_A - \beta_B)} \quad (38)$$

For source particle that produce interaction potential in a point on probe particle on write (35) as:

$$I = A \cdot e^{C\frac{R_s}{R} + 4p^2 B\left(\vec{d} + \vec{\Delta}\right)^2} = A \cdot e^{4p^2 B\Delta^2} \cdot e^{C\frac{R_s}{R} + 4p^2 Bd^2 + 8p^2 B\Delta d \cos\alpha} \quad (39)$$

where $\alpha$ is angle between $\delta$ and $\Delta$. If distant R between the two particles is very large, then on can work in first order approximation and (39) become:

$$I \cong A \cdot e^{4p^2 B\Delta^2} \cdot \left(1 + C\frac{R_s}{R} + 4p^2 Bd^2 + 8p^2 B\Delta d \cos\alpha\right) \quad (40)$$

Considering that the two particles have the same charge, let's calculate the quantity:

$$a = \frac{I_B - I_A}{\overline{AB}} \cong \frac{A \cdot e^{4p^2 B\Delta^2}}{\overline{AB}} \cdot \left[CR_s\left(\frac{1}{R_B} - \frac{1}{R_A}\right) - 4p^2 Bd_A^2 - 8p^2 B\Delta d_A \cos\alpha_A\right] \quad (41)$$

Because:

$$R_B = R + r_0 \cos \mathbf{a}_B$$
$$R_A = R + r_0 \cos \mathbf{a}_A$$
$$R \gg \overline{AB}$$
$$\mathbf{d} = \Delta \frac{R_0^2}{R^2}$$
(42)

By introducing notation $A_0 = A\exp(4\pi^2 B\Delta^2)$ and using expression for $\delta$ from (35), after some calculations (and neglecting the superior order terms) on obtain:

$$a \cong A_0 CR_s \frac{1}{R^2} \frac{r_0(\cos \mathbf{a}_A - \cos \mathbf{a}_B)}{\overline{AB}} - 8\mathbf{p}^2 A_0 B\Delta \mathbf{d} \frac{\cos \mathbf{a}_A}{\overline{AB}} - 4\mathbf{p}^2 A_0 B\Delta \cos \mathbf{a}_A \frac{2r_0 \cos \mathbf{a}_A}{\overline{AB}} \cdot \left(-\frac{2\Delta R_0^2}{R^3}\right) - \ldots$$
$$- 2\mathbf{p}^2 A_0 B \frac{2r_0 \cos \mathbf{a}_A}{\overline{AB}} \left(-\frac{4\Delta^2 R_0^4}{R^5}\right)$$
(43)

Relationship (43) can be writed:

$$a \cong A_0 CR_s \frac{1}{R^2} \frac{r_0(\cos \mathbf{a}_A - \cos \mathbf{a}_B)}{\overline{AB}} - 8\mathbf{p}^2 A_0 B\Delta \mathbf{d} \frac{\cos \mathbf{a}_A}{\overline{AB}} - 4\mathbf{p}^2 A_0 B\Delta \cos \mathbf{a}_A \frac{2r_0 \cos \mathbf{a}_A}{\overline{AB}} \cdot \frac{d\mathbf{d}}{dR} - \ldots$$
$$- 2\mathbf{p}^2 A_0 B \frac{2r_0 \cos \mathbf{a}_A}{\overline{AB}} \cdot \frac{d(\mathbf{d}^2)}{dR}$$
(44)

The last two terms of right member of (44) depends on $1/R$ by powers greater than 2 and, for instant, they can be neglected. Further, on see the consequences of those terms.

For interaction between opposite charge particles the relationship (44) on write:

$$a \cong A_0 CR_s \frac{1}{R^2} \frac{r_0(\cos \mathbf{a}_A - \cos \mathbf{a}_B)}{\overline{AB}} + 8\mathbf{p}^2 A_0 B\Delta \mathbf{d} \frac{\cos \mathbf{a}_B}{\overline{AB}} + 4\mathbf{p}^2 A_0 B\Delta \cos \mathbf{a}_B \frac{2r_0 \cos \mathbf{a}_B}{\overline{AB}} \cdot \frac{d\mathbf{d}}{dR} + \ldots$$
$$+ 2\mathbf{p}^2 A_0 B \frac{2r_0 \cos \mathbf{a}_B}{\overline{AB}} \cdot \frac{d(\mathbf{d}^2)}{dR}$$
(45)

Let's focusing now on (44) with only the first two terms in right member:

$$a \cong A_0 CR_s \frac{1}{R^2} \frac{r_0(\cos \mathbf{a}_A - \cos \mathbf{a}_B)}{\overline{AB}} - 8\mathbf{p}^2 A_0 B\Delta \mathbf{d} \frac{\cos \mathbf{a}_A}{\overline{AB}}$$
(46)

On observe that first term is due to difference between concentration of fluctuations that source produces on the two hemispheres of probe (→ gradient of concentration). The second term is due to difference between resultant of asymmetry of fluctuations distribution on the two hemispheres of probe (→ gradient of asymmetry).

Let's introduce the distance AB from (38) in the second term of (46):

$$a \cong A_0 CR_s \frac{r_0(\cos \mathbf{a}_A - \cos \mathbf{a}_B)}{\overline{AB}} \frac{1}{R^2} - \frac{8\mathbf{p}^2 A_0 B\Delta \mathbf{d}}{\sqrt{2} \cdot r_0} \frac{\cos \mathbf{a}_A}{\sqrt{1 - \cos \mathbf{a}_A \cos \mathbf{a}_B - \sin \mathbf{a}_A \sin \mathbf{a}_B \cos(\mathbf{b}_A - \mathbf{b}_B)}}$$
(47)

The maximum for:

$$\frac{\cos \mathbf{a}_A}{\sqrt{1 - \cos \mathbf{a}_A \cos \mathbf{a}_B - \sin \mathbf{a}_A \sin \mathbf{a}_B \cos(\mathbf{b}_A - \mathbf{b}_B)}}$$
(48)

correspond to AB parallel to Ox, which is normal because difference between resultant of asymmetry of the two hemispheres must be maximum on Ox direction (the direction of the two interacting particles). Also: $AB = 2r_0 \cos \alpha_B$.

On results:

$$a \cong -A_0 CR_s \frac{1}{R^2} + \frac{4\mathbf{p}^2 A_0 B\Delta \mathbf{d}(R)}{r_0} = -A_0 CR_s \frac{1}{R^2} + \frac{4\mathbf{p}^2 A_0 B\Delta^2 R_0^2}{r_0} \frac{1}{R^2}$$
(49)

Following the same demonstration, for the opposite charge particles on obtain the relationship (49) too. The relationships (44) and (45) can be combined in a single relationship:

$$a \cong -A_0 C R_s \frac{1}{R^2} + \frac{4p^2 A_0 B \Delta d(R)}{r_0} \pm 4p^2 A_0 B \Delta \overline{\cos a_B} \cdot \frac{dd(R)}{dR} + 2p^2 A_0 B \cdot \frac{d(d^2(R))}{dR} \quad (50)$$

The sign of third term is (-) when interacting particles have charges of the same sign and (+) when they are of opposite sign.

We'll use (50) later, when we'll see the significance of the last two terms.

### 3.1. Gravitational interaction and electrostatic interaction

Let source-probe distance be R. On compare the probe acceleration due to gradient of concentration and asymmetry of fluctuation distribution (relationship (49)) with probe acceleration (charge $e_m$ and mass $m_0$) due to gravitational and electrostatic fields of source (charge $e_M$ and mass M):

$$a = -GM \frac{1}{R^2} + \frac{e_M e_m}{4p e_0} \frac{1}{m_0} \frac{1}{R^2} \quad (49')$$

By comparing (49) with (49') on observe firstly that acceleration depends on distance is the same. Secondly, on observe that first term in both relationships is attractive and the second can be both attractive (opposite orientations for δ and Δ in (49) – opposite charges in (49')) and repulsive (same orientations for δ and Δ in (49) – same sign charges in (49')). First term of (49') is in proportion with source mass and first term of (49) is in proportion with source radius. Second term of (49') is in inverse proportion with probe mass and second term of (49) is in inverse proportion with probe radius. On conclude that (49) and (49') are the same. On identify:

$$A_0 C R_s = GM \quad (51)$$

On observe that it is proportion between mass and radius of particles. As result, on can write a similar relationship between probe mass and radius:

$$A_0 C r_0 = G m_0 \quad (52)$$

By introducing (52) in (49) results:

$$a \cong -A_0 C R_s \frac{1}{R^2} + \frac{4p^2 A_0^2 B C \Delta^2 R_0^2}{G} \frac{1}{m_0} \frac{1}{R^2} \quad (53)$$

Allowing (49') it results:

$$\frac{4p^2 A_0^2 B C \Delta^2 R_0^2}{G} = \frac{e_M e_m}{4p e_0} \quad (54)$$

As results, the movement of probe due to gravitational attraction of source is in fact due to gradient of concentration of temporal fluctuations. This gradient produces the gradient of average lifetime of fluctuations.

The electrostatic interaction is due to difference between asymmetry of distribution of fluctuations on the two hemispheres of probe. This difference is due to vectorial addition of asymmetry produces by source with asymmetry at probe surface.

Of course, the probe acts alike on source.

### 3.2. Gravitational mass, inertial mass, inertia

From above it is obvious that probe acceleration due to difference in asymmetry of fluctuation distribution must be in inverse proportion with its radius, as it results from (49). Considering the particle as source of gradient of fluctuation concentration (*gravitational field*), on see that field strength is in proportion with source radius. On the other hand, inertial mass that enters in calculus of acceleration produces by difference of asymmetry of fluctuation distribution (*electrostatic field*) is in proportion with particle's radius too. As result, both gravitational and inertial masses can be replaced by particle's radius. In this way, the equality between the two types of masses (the equivalence principle in GRT) it is naturally satisfied. Much more, gravitational and inertial masses are identical because are physical quantities resulting from the same property: particle's radius.

On shortly, the inverse proportionality between electrostatic part of acceleration of particle and its mass is due to difference between the two hemispheres asymmetry of distribution is the same whatever particle's radius is. The particles with big radius (i.e. big mass) accelerate slower than particles with small radius.

On observe that from Temporal Fluctuation Model naturally results that acceleration due to gradient of fluctuations concentration does not depend on probe radius ($\rightarrow$ gravitational interaction) and acceleration due to difference of asymmetry of fluctuations distribution is in inverse proportion with probe radius ($\rightarrow$ electrostatic interaction).

As result, the model allow to understanding the very nature of inertia and its relation with gravity, without a supplemental postulate – equivalence principle.

### 3.3. Elementary electric charge

From (54) on deduce the elementary electric charge:

$$e = 4pA_0\Delta R_0 \sqrt{\frac{p\varepsilon_0 BC}{G}} \tag{55}$$

Using a unit system with $4\pi\varepsilon_0=1$, on can write the elementary electric charge as:

$$e = 2pA_0\Delta R_0 \sqrt{\frac{BC}{G}} \tag{56}$$

Because for every charged particle the asymmetry on its surface is the same (with the maximum value $\Delta$) and the rest of quantities from (55) are constant, result that e is constant and its sign is the sign of $\Delta$. Results too that the absolute value for charge of particles is the same.

### 3.4. Asymmetry of fluctuations distribution and electric field strength

The modulus of probe (e,$m_0$) acceleration in electrostatic field with strength **E** that is produce by source is:

$$a = \frac{eE}{m_0} \tag{57}$$

By compare with the acceleration that difference of asymmetry produces (the second term of (53)) on write:

$$E(R) = \frac{4p^2 A_0^2 BC\Delta}{eG} \cdot d(R) \tag{58}$$

where on use definition from (35) for d at distance R. E(R) and $\delta$(R) have the same variation with distance R If $\Delta$ and e are the absolute values of asymmetry maximum and elementary charge, on can use the same sign convention, both for E and $\delta$. On write the vectorial form of relationship (58):

$$\vec{E}(R) = \frac{4p^2 A_0^2 BC\Delta}{eG} \cdot d(R) \frac{\vec{R}}{R} \tag{59}$$

In TFM the quantity $\delta \sim R^{-2}$ show that asymmetry distribution decrease with distance. The electric field strength shows that electric field decrease with $R^2$ too. Remembering that superposition is produced in the same way for **E** and for $\delta$, it is an inevitable conclusion that the physical entity named electric field is, in fact, the field of asymmetry of distribution for temporal fluctuation orientation.

### 3.5. The principle of action and reaction. Force

Let' take the same two particles on Ox axis at distance R one each other. Source has radius $R_s$ and probe have radius $r_0$. On notate $n=R_s/r_0$. Let's consider for the beginning that particles interact only gravitational. Each particle is subject of influence from the other particle that produces its own gradient of interaction potential. The directions of those gradients are parallel with Ox. Using the first order approximation, on write:

$$\begin{aligned} \left(\frac{\Delta I}{\Delta R}\right)_{sursa} &= -A_0 C \frac{R_s}{R} \\ \left(\frac{\Delta I}{\Delta R}\right)_{proba} &= -A_0 C \frac{r_0}{R} \end{aligned} \tag{60}$$

Source produces a n times greater gradient on probe than the probe produces on source. Because probe radius in n times smaller than source radius, results that radius of one particle multiplied by gradient produces by the other particle is the same for the two particles:

$$r_0\left(\frac{\Delta I}{\Delta R}\right)_{sursa} = R_s\left(\frac{\Delta I}{\Delta R}\right)_{proba} \tag{61}$$

On introduce (60), (51) and (52) in (61). It results:

$$\frac{GMm_0}{R^2} = \frac{Gm_0 M}{R^2} \quad \text{that is} \quad F_{sursa \to proba} = F_{proba \to sursa} \tag{62}$$

On conclude that source act on probe same as probe act on source, which is exactly the principle of action and reaction. On could say that action and reaction originate in that the every particle produces the same type of action one each other. In the above demonstration every particle produces a gradient of fluctuations concentration in the region where is the other particle.

On observe that difference between interaction potential produces by source particle between two points A and B situated on the surface of the probe particle (AB is parallel with direction of maximum gradient) is exactly the force that source act on probe particle:

$$\overrightarrow{F_{AB}} = (I_B - I_A)\frac{\overrightarrow{AB}}{\overline{AB}} \tag{63}$$

Following a similar reasoning, if the particles interact due to asymmetry of distribution of fluctuation orientation, on arrive to the same conclusion.

Because all interactions that determine mechanical effects on bodies are based on electrical and gravitational interactions, the principle of action and reaction have a general validity.

Let's see an important consequence of fact that principle of action and reaction is due to the actions of the same kind. If S produces only gradient of concentration ($\to$ gravitational field) and P produces only gradient of asymmetry ($\to$ electric field), the action of S on P differ of action of P on S that is the action-reaction principle is violated. On can imagine even closed systems that accelerate by themselves.

As result, the Newton's third law must reformulate as follows:

***For an isolated system, for the bodies between exert interactions of the same kind, the action and reaction are equal and opposite.***

A direct consequence of Newton's third law reformulation is the necessity of momentum law conservation reformulation, as follows:

***For an isolated system, for the bodies between exert interactions of the same kind, the resulting momentum is zero.***

We'll see that in some circumstances, the supplemental terms from (50) for acceleration will lead to Newton's third law and momentum law violation.

**3.6. Unified potential. Gravitational potential. Electric potential**

Let it be the source S that produces a concentration gradient and asymmetry of distribution for fluctuations. The interaction potential at distance R, if in that point is a particle, is given by relationship:

$$I(R) = A \cdot e^{C\frac{R_s}{R} + 4p^2 B \Delta^2 \left(\frac{R_0^2}{R^2}\right)^2} \tag{64}$$

On assume that at infinite the potential is null. On define the new potential as:

$$\Phi_I(R) = A - I(R) \tag{65}$$

which in the weak field approximation on write:

$$\Phi_I(R) \cong -\left[AC\frac{R_s}{R} + 4\mathbf{p}^2 AB\Delta^2\left(\frac{R_0^2}{R^2}\right)^2\right] \tag{66}$$

It can define the field strength induced by interaction potential:

$$\vec{i} = -\text{grad}\,\Phi_I(R) \tag{67}$$

where on use the usual definition for field, which derive from potential.
For a source with spherical symmetry, on first order approximation, the intensity become:

$$\vec{i} = -A\left(CR_s\frac{1}{R^2} + 4\mathbf{p}^2 B\Delta^2 R_0^4 \frac{4}{R^5}\right)\frac{\vec{R}}{R} = \left[-ACR_s\frac{1}{R^2} + 4\mathbf{p}^2 AB \frac{d(\mathbf{d}^2(R))}{dR}\right]\frac{\vec{R}}{R} \tag{68}$$

Let's write the probe acceleration due to source field, as it results from relationship (50) which is write in vectorial notation for source with spherical symmetry:

$$\vec{a} \cong \left[-A_0 CR_s\frac{1}{R^2} + \frac{4\mathbf{p}^2 A_0 B\Delta \mathbf{d}(R)}{r_0} \pm 4\mathbf{p}^2 A_0 B\Delta\overline{\cos\mathbf{a}_B} \cdot \frac{d\mathbf{d}(R)}{dR} + 2\mathbf{p}^2 A_0 B \cdot \frac{d(\mathbf{d}^2(R))}{dR}\right]\frac{\vec{R}}{R} \tag{69}$$

Because for every central force the work function is independent by way and on can define a potential [12], on consider that the acceleration given by (69) can derive from potential:

$$\Phi(R) \cong -\left[A_0 CR_s\frac{1}{R} - \frac{4\mathbf{p}^2 A_0 B\Delta^2 R_0^2}{r_0}\frac{1}{R} \pm 4\mathbf{p}^2 A_0 B\Delta\overline{\cos\mathbf{a}_B}\mathbf{d}(R) + 2\mathbf{p}^2 A_0 B\mathbf{d}^2(R)\right] \tag{70}$$

Which it obtained by acceleration integration using the condition that the potential is zero at infinity. On nominate unified potential the quantity Φ(R).
Let's compare the acceleration given by (69) with field strength determined by interaction potential (68). On observe that the second and third terms of (69) due to charge of probe (Δ≠0). Secondly, on observe that the rest of the terms of **a** are like the terms from (68) and they are attractive. First term is due to the gradient of concentration and the second is due to the gradient of asymmetry of distribution for fluctuations orientation. On observe that the presence of probe modify the strength of interaction because the constant A from (68) is replaced by $A_0$ in (69). The same discussion is for $\Phi_I$ and Φ.
Let's focusing on the first two terms of (70) which are the most important. Using (52) and (54) it results:

$$\Phi(R) \cong -\frac{GM}{R} + \frac{e_m}{m_0}\frac{e_M}{4\mathbf{p}e_0}\frac{1}{R} \tag{71}$$

On observe that the first term in gravitational potential of source at distance R, and the second is the electrostatic potential of source at distance R multiplied by probe's charge to mass ratio.
On apply the gradient with (-) sign on Φ(R) and the result is the acceleration of probe in gravitational and electrostatic fields:

$$\vec{a} = -\text{grad}\left(\Phi_{grav} + \frac{e_m}{m_0}\Phi_{el}\right) = \vec{g} + \frac{e_m}{m_0}\vec{E} \tag{72}$$

where **g** and **E** are the gravitational field strength and electrostatic field strength that the source produces.
As conclusion, on see that using the Temporal Fluctuations Model results the gravitational and electrostatic potentials. On observe too, that the probe influences the potential. This is easy to see if on compare the gravitational potential from Φ(R) expression (probe present) with the correspondent term from $\Phi_I(R)$ expression (without probe).
It must observe that at unified potential expression (70) contribute the usual convention that the potential at infinity is null. Otherwise, the potential at infinity is $A_0$.
The last two terms from (70) will be subject of future discussion.

## 4. Metric tensor

In this section on find the Schwarzschild metric tensor elements by starting from TFM. On find the main conclusion of GRT without notion of curved spaces.

Let it be a source that produces only gradient of fluctuations concentration ($\rightarrow$ gravitational field). The whole energy of probe at distance R in the field of source is [24]:

$$E = m_0 c^2 + m_0 \Phi_{grav} = m_0 c^2 \left(1 - \frac{GM}{c^2 R}\right) \tag{73}$$

Let it be an atom as probe. The difference between two atomic energy levels is $\Delta E_0 = E_{02} - E_{01}$ outside the gravitational field. In presence of gravitational field of source, the energy of the two atomic levels and the difference between them become:

$$E_1 = E_{01}\left(1 + \frac{\Phi_{grav}}{c^2}\right)$$
$$E_2 = E_{02}\left(1 + \frac{\Phi_{grav}}{c^2}\right) \tag{74}$$
$$\Delta E = \Delta E_0 \left(1 + \frac{\Phi_{grav}}{c^2}\right)$$

On divide last relationship of (74) by Planck constant h and on obtain the link between frequency of the photons emitted in desexcitation process:

$$\boldsymbol{\nu} = \boldsymbol{\nu}_0 \left(1 + \frac{\Phi_{grav}}{c^2}\right) \tag{75}$$

where $\nu$ is the frequency of the emitted photon by the atom situated in field and $\nu_0$ is the frequency of the emitted photon by the atom outside the field. Using the duration (inverse of frequency) on write:

$$T = \frac{T_0}{\left(1 + \frac{\Phi_{grav}}{c^2}\right)} \cong T_0 \left(1 - \frac{\Phi_{grav}}{c^2}\right) = T_0\left(1 + \frac{GM}{c^2 R}\right) \tag{76}$$

The above relationship may be generalized as follows: the duration of any event is minimum in the absence of gravitation and increase when approach to gravitational field source.

The relationships (75) and (76) explain the result of Pound-Rebka experience by considering that difference between two atomic levels is smaller for the atom that is closer to gravitational field source, instead of photons red shift. The above interpretation is sustained by measurements with airborne atomic clocks and by the corrections needed for the GPS system.

Let's treat the same problem from TFM viewpoint. On consider that source produces only gradient of fluctuations concentration so in the relationships (64), (66) and (70) remain only the appropriate term. Because average fluctuation's lifetime influences the duration of events and the lifetime is in proportion with interaction potential, on can write:

$$T_0 \sim I_0 = A$$
$$T \sim I = A e^{C \frac{R_s}{R}} \qquad \Rightarrow \quad T = T_0 e^{C \frac{R_s}{R}} \cong T_0 \left(1 + C \frac{R_s}{R}\right) \tag{77}$$

Using the link between mass and radius of particles (51) on write:

$$T \cong T_0 \left(1 + \frac{GM}{A_0 R}\right) \tag{78}$$

Relationships (76) and (78) are the same if $A_0=c^2$. On see further the significance of this equality.

Demonstration from (77) is not valid for $\Phi_I$ and $\Phi$ because those potentials use the convention that at infinity the potential is zero. If integration of acceleration for obtaining the potential from (70) with condition that the potential is $A_0$ at infinity, then the demonstration from (77) with $\Phi$ lead to the same result concerning the ratio of duration.

On conclude that vacuum (or space-time continuum) determine in every point of it the potential:

$$A_0 = c^2 \tag{79}$$

and relationship $E=mc^2$ signify that the particle with mass m have potential energy E with reference to vacuum.

***Any particle with mass m have the potential $c^2$ and potential energy $E=mc^2$ with reference to vacuum (space-time continuum)***

The vacuum potential $c^2$ is linked with quantity c, which represents the fluctuations velocity. In the same time c is the velocity of interactions between fluctuations that represents the light velocity far away from any source of fields. On specify this because on will see that light velocity modify in the presence of fields.

Another consequence is the fact that the decrease of energy for a particle that came from infinity to vicinity of a gravitational field source is due to its mass decrease. This fact could be interpreted that between probe and source appear a binding energy, similar to what is happened at formation of nucleus from nucleons.

On notate with k the factor by which on express the mass decrease when bring the probe in gravitational field. If $m_0$ is the probe mass at infinity and m(R) is the mass at distance R from source, on write:

$$m=m_0/k \tag{80}$$

For particle in gravitational field on write Lagrange equation for free particle [19] in which on replace $m_0$ by $m=m_0/k$:

$$\frac{d}{dt}\left(\frac{m_0}{k}\frac{\vec{v}}{\sqrt{1-v^2/c^2}}\right) = \nabla\left(-\frac{m_0}{k}c^2\sqrt{1-v^2/c^2}\right) \tag{81}$$

On work in no relativistic case, spherical symmetry and stationary case. It results the equation:

$$\frac{1}{k}\frac{dv}{dt} = -c^2 \frac{d}{dR}\left(\frac{1}{k}\right) \tag{82}$$

Because the body acceleration in gravitational field is:

$$\frac{dv}{dt} = -\frac{GM}{R} \tag{83}$$

It results the equation:

$$\frac{dk}{k} = -\frac{GM}{c^2}\frac{dR}{R^2} \tag{84}$$

Using the condition $R\to\infty \implies k\to 1$, the solution is:

$$k = e^{\frac{GM}{c^2 R}} = e^{-\frac{\Phi_{grav}}{c^2}} \tag{85}$$

Which in the weak field approximation become:

$$k \cong 1 + \frac{GM}{c^2 R} = 1 - \frac{\Phi_{grav}}{c^2} \tag{86}$$

On observe that k from the above relationship is the same with the k from (73) – (78) relationships. If for k on use the expression from (85), the relationships (73) – (78) can be write not only for the weak field approximation.

Because in TFM the particle's mass is in proportion with its radius, by using (80) on find the link between probe's radius in absence of field ($r_0$) and the same radius in presence of field (r):

$$r = \frac{r_0}{e^{-\frac{\Phi_{grav}}{c^2}}} \tag{87}$$

As conclusion, the energy, mass and distances decrease in the presence of gravitational field and duration of events increases.

Let's see the consequences of relationship (87) for the two atomic levels. Let $r_{0s}$ and $r_{0i}$ the radius for the superior and inferior atomic levels outside field. Let $E_{0s}$ and $E_{0i}$ the appropriate energies. On notate with $r_s$, $r_i$, $E_s$ and $E_i$ the same quantities for the atom in gravitational field. Remembering the charge conservation on write:

$$E_{0s} = m_0 c^2 - \frac{e}{4\pi\varepsilon_{00} r_{0s}}$$

$$\Rightarrow \quad \Delta E_0 = -\frac{e}{4\pi\varepsilon_{00}}\left(\frac{1}{r_{0s}} - \frac{1}{r_{0i}}\right)$$

$$E_{0i} = m_0 c^2 - \frac{e}{4\pi\varepsilon_{00} r_{0i}} \tag{88}$$

$$corresponding \qquad \Delta E = -\frac{e}{4\pi\varepsilon_0}\left(\frac{1}{r_s} - \frac{1}{r_i}\right)$$

where $\varepsilon_{00}$ and $\varepsilon_0$ are the electrical permittivities at infinity and respectively in field. Using (87) for distances and $\Delta E = \Delta E_0/k$ for energy results:

$$\varepsilon_0 = \varepsilon_{00} k^2 \tag{89}$$

The studies on fundamental constants in various cosmological conditions impose that fine structure constant α must be constant [35]. The conservation of angular momentum of circularly polarized photon in vacuum impose that $\hbar$ must be constant. On write for fine structure constant in both situations:

$$\alpha_0 = \frac{e^2}{4\pi\varepsilon_{00}\hbar c} = \frac{e^2}{4\pi\hbar}\sqrt{\frac{\mu_{00}}{\varepsilon_{00}}}$$

$$\alpha = \frac{e^2}{4\pi\hbar}\sqrt{\frac{\mu_0}{\varepsilon_0}} \tag{90}$$

Because $\alpha = \alpha_0$ it must that $\mu_{00} = k'\mu_0$ and $\varepsilon_{00} = k'\varepsilon_0$. Using (89) results $k' = k^2$. If on notate the light speed in the zone with gravitational field by $c_c$ it results:

$$c_c = \frac{c}{k^2} \tag{91}$$

As result, the increase of gravitational field produce the decrease of light speed that become zero when k→∞ (for R→0). If on use for k the expression for weak field (86), the light speed become zero for R equals Schwarzschild radius. Because the complete expression for k is (85), then the Schwarzschild radius looses its significance.

The infinitesimal space-time interval for a point at infinity is:

$$ds^2 = c^2 dt_0^2 - \left(dx_0^2 + dy_0^2 + dz_0^2\right) \tag{92}$$

In a region with k>1 the rods for length measurements shrink k times and duration of events dilate k times:

$$dt = kdt_0; \quad dx = dx_0 / k; \quad dy = dy_0 / k; \quad dz = dz_0 / k \tag{93}$$

On introduce (93) in (92). If on use local quantities for the infinitesimal space-time interval, it results:

$$ds^2 = \frac{1}{k^2} c^2 dt^2 - k^2 \left( dx^2 + dy^2 + dz^2 \right) \tag{94}$$

The above expression for ds can be compared with tensorial expression for the infinitesimal space-time interval from GRT:

$$ds^2 = g_{ij} dx^i dx^j \tag{95}$$

where $g_{ij}$ define the metric tensor. By comparing (94) with (95) on identify:

$$dx^0 = cdt; \quad g_{00} = \frac{1}{k^2}; \quad g_{11} = g_{22} = g_{33} = -k^2; \quad g_{ij} = 0 \;\; for \;\; i \neq j \tag{96}$$

In weak field approximation, at distance R from gravitational field source:

$$k^2 \cong 1 + 2\frac{GM}{c^2 R} \tag{97}$$

With $k^2$ from (97), the quantity $g_{ij}$ from (96) define exactly the Schwarzschild metrics.

On found that starting from TFM on retrieve the elements of Schwarzschild metric tensor without the necessity of curved space. In this way TFM retrieve the main consequences of GRT: gravitational red shift, light ray bending and perihelion advance, all this in the frame of Euclidean geometry.

Relationship (94) is valid for light if on make ds=0. If on consider that light propagates along Ox, the light speed in gravitational field measured by an observer positioned outside field is:

$$\frac{dx}{dt} = c_c = \frac{c}{k^2} \tag{98}$$

that is identical to (91), which is obtained using another way.

It is natural to consider that if the source produce not only gravitational field but also electrostatic field, on can replace the gravitational potential by unified potential (71) in k expressions from (85) and (86). For a better approximation on can use (70) instead (70) for unified potential. Relationships (85) and (86) become:

$$k = e^{-\frac{\Phi(R)}{c^2}} \cong 1 - \frac{\Phi(R)}{c^2} \tag{99}$$

All the results and conclusions founded in this section are valid for the new k. On observe that the presence of electrostatic field produces modifications on the same physical quantities as the gravitational field. On must emphasize that in electrostatic (or electric) field there is the possibility that k<1.

On rewrite the term from weak field approximation from (99) as:

$$k \cong \frac{c^2 - \Phi(R)}{c^2} \tag{100}$$

The denominator is the vacuum potential and the numerator is the unified potential if the potential at infinity equals vacuum potential. As result on can consider that presence of particles influence the value of vacuum potential and k show the amount of that influence. On can say that k factor measure the degree of *vacuum polarization* influenced by gradient of concentration and/or asymmetry of distribution of fluctuations.

Following the model offered by (64) for the interaction potential (with convention that at infinity I→ A) on enunciate next conjecture:

If the value of unified potential approach to vacuum potential when distance become infinity, the unified potential is given by relationship:

$$\Phi_u(R) = c^2 e^{-\frac{\Phi(R)}{c^2}} \tag{101}$$

where $\Phi(R)$ is given by (70). On see that in weak field approximation:

$$\Phi_u(R) \cong c^2 - \frac{\Phi(R)}{c^2} \tag{102}$$

It is possible that in the frame of a much evolved TFM, in which the function of distribution of fluctuations orientation have a more adequate shape and using superior order approximation of interaction potential in demonstration for acceleration in section 3, the above conjecture can be demonstrated.

If on consider true the relationship (101) and on notate the vacuum potential by $\Phi_v$, then k will be give by the expression:

$$k = \frac{\Phi_u(R)}{\Phi_v} \tag{103}$$

and interpretation of k as degree of vacuum polarization it is not limited to weak field approximation.

On remember that the effect of source can be seen only by its action on probe. If on refer to (64), (101) and (102) because $I(R\rightarrow\infty)\rightarrow A$ and $\Phi_u(R\rightarrow\infty)\rightarrow A_0$ on can consider that vacuum potential related to itself is A and the presence of one particle modify it to $A_0=c^2$. The above discussion may be interpreted that observer (probe) influences the value of measured quantity (the potential).

### 5. Electrodynamics and gravitodynamics

#### 5.1. Unified potential of source in movement

For the porpoises of the next discussion on simplify the notation of unified potential from (71), as follows:

$$\Phi(R) = \frac{\Phi_0}{R} \quad \text{where} \quad \Phi_0 = -GM + \frac{e_m}{m_0}\frac{e_M}{4\pi\varepsilon_0} \tag{104}$$

Let a source in movement with velocity **v** along Ox axis. Let's find the unified potential produced by source in a point P at great distance (figure 6):

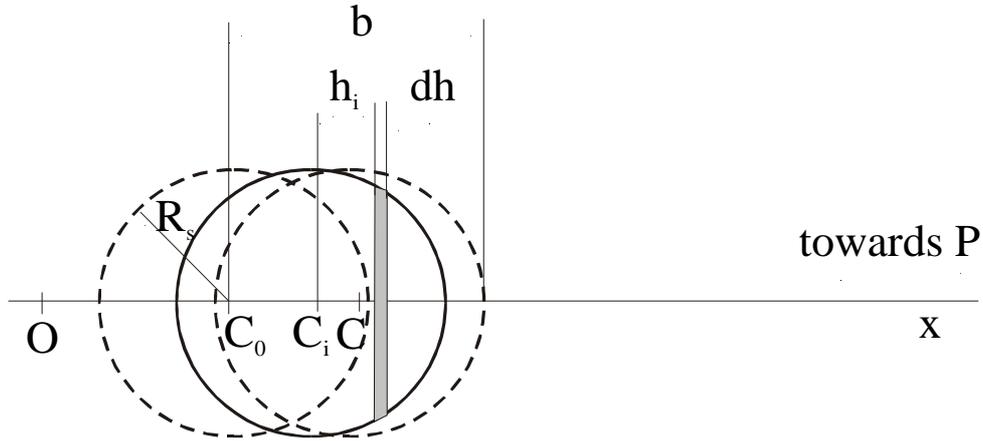

Figure 6

On consider that right hemisphere of source is divided in surface elements dS with infinitesimal thickens dh, each of them generating the potential dΦ. On obtain the potential in P by integrate dΦ. Because the source moves, for finding the potential at the same moment it must integrate dΦ at different positions. The three source position from figure 6 correspond to following moments: integration beginning ($C_0$), moment i ($C_i$) and the end of integration (C). If $C_0P$ is very large on consider that distant between each surface element and P is the same – R. It results:

$$d\Phi = \frac{\Phi_0}{R}\frac{2\pi R_s dh}{2\pi R_s^2} \quad \Rightarrow \quad \Phi = \int_0^b \frac{\Phi_0}{R}\frac{dh}{R_s} = \frac{\Phi_0}{R}\frac{b}{R_s} \tag{105}$$

b is Rs plus distance that source moves in duration:

$$\Delta t = (t - R_1/c) - (t - R_N/c) = b/c \tag{106}$$

where the potential in P is calculate at instant t, $R_1$ is the distance between surface element at the beginning of integration and $R_N$ is the same distance but for the last element of integration. It results:

$$b = \frac{a}{1 - v/c} \tag{107}$$

The unified potential in P is:

$$\Phi = \frac{\Phi_0}{R} \frac{1}{1 - v/c} \tag{108}$$

If v have another orientation than R, then:

$$\Phi = \frac{\Phi_0}{\left[R - \left(\vec{v} \cdot \vec{R}/c\right)\right]_{ret}} \tag{109}$$

where all quantities from denominator are calculated at retarded instant $t - R_{ret}/c$.

Like in electrodynamics on define the vector potential:

$$\vec{A} = \Phi \frac{\vec{v}}{c^2} \tag{110}$$

Introducing (109) in (110) it results:

$$\vec{A} = \frac{\Phi_0}{\left[R - \left(\vec{v} \cdot \vec{R}/c\right)\right]_{ret}} \cdot \frac{\vec{v}}{c^2} \tag{111}$$

Relationship (109) and (111) are similar with relationships for Lienard-Wiechert potentials from electrodynamics.

If source generate only gradient of fluctuations concentration, then $\Phi_0$ reduces to (-GM) and aforesaid relationships describe gravitational scalar and vector potentials. If source generate only asymmetry of fluctuations distribution, then aforesaid relationships describe Lienard-Wiechert potentials up to a constant factor – the probe charge-mass ratio.

At this moment on can show that TFM can justify the existence of vector potential. Let's illustrate for electrodynamics: if the source motion is linear and monotonous, the fluctuations ensemble around source moves together with the source. This is compatible with the fact that electric field travel together with his source, if the motion is linear and monotonous [30], [35]. If source changes the motion characteristics, the field conserves the initial motion.

Getting back to TFM, the fluctuations ensemble drift motion (velocity **v**) is equivalent with production of a supplemental potential. This potential must be in proportion with electrostatic potential and v/c. Because is a quantity linked with motion, its temporal variation must produce an effect which is similar to the effect that gradient of electrostatic potential produces. Therefore, it is a contribution to the electric field strength part, which is parallel to velocity:

$$-\frac{1}{c} \frac{\partial}{\partial t} \left( \frac{\vec{v}}{c} \Phi_{electrostatic} \right) \tag{112}$$

where on introduce 1/c because of dimensional coherence and (-) sign for the coherence of sign conventions. As result the electric field strength become:

$$\vec{E} = -\nabla \Phi_{el} - \frac{\partial}{\partial t}\left(\frac{\vec{v}}{c^2}\Phi_{el}\right) \tag{113}$$

where the quantity between brackets is exactly the vector potential.
In the same manner on justify the existence of gravitational vector potential.

On get back to unified scalar and vector potentials. By analogy with electrodynamics on can define the unified quadripotential as follows:

$$A_m = \left[\frac{1}{c}\Phi, \vec{A}\right] \qquad m = 0,1,2,3 \tag{114}$$

If source moves with velocity v, the unified potential measured in rest reference frame is:

$$\Phi' = \frac{\Phi - \vec{v}\cdot\vec{A}}{\sqrt{1-v^2/c^2}} \qquad \vec{A}' = \frac{\vec{A} - \vec{v}/c^2 \cdot \Phi}{\sqrt{1-v^2/c^2}} \tag{115}$$

In the same reference frame the k factor from section 4 is:

$$k' \cong \frac{c^2 - \Phi'(R)}{c^2} \tag{116}$$

Again, by analogy with electrodynamics on define the strength and magnetic induction for unified field:

$$\vec{E}_u = -\nabla\Phi - \frac{\partial\vec{A}}{\partial t} \qquad \vec{B}_u = \nabla \times \vec{A} \tag{117}$$

By analogy with Maxwell equations from electrodynamics on can write:

$$\nabla \cdot \vec{E}_u = \frac{\rho_\Phi}{\varepsilon_\Phi}$$

$$\nabla \cdot \vec{B}_u = 0$$

$$\nabla \times \vec{E}_u = -\frac{\partial \vec{B}}{\partial t} \tag{118}$$

$$c^2 \cdot \nabla \times \vec{B}_u = \frac{\vec{j}_\Phi}{\varepsilon_\Phi} + \frac{\partial \vec{E}}{\partial t}$$

Where $\rho_\Phi$ is *unified charge density* and $\varepsilon_\Phi$ is *unified vacuum permittivity*:

$$\frac{\rho_\Phi}{\varepsilon_\Phi} = -4\pi G \rho_M + \frac{\rho_{eM}}{\varepsilon_0}\frac{\rho_{em}}{\rho_m} \tag{119}$$

$\rho_{eM}$ and $\rho_{em}$ are the electrical charge density for source and respectively for probe
$\rho_M$ and $\rho_m$ are the mass density for source and respectively for probe

As well, the *unified current density*:

$$\vec{j}_\Phi = \rho_\Phi \vec{v} \tag{120}$$

Because unified potential have the same dimension as gravitational potential, on can (formally) identify $\rho_\Phi = \rho_M$ in (119). On write in this situation:

$$\frac{1}{\varepsilon_\Phi} = -4\pi G + \frac{1}{\varepsilon_0}\frac{\rho_{eM}}{\rho_M}\frac{\rho_{em}}{\rho_m} \tag{121}$$

By dissociation of the two contributions (gravitational and electrical), from (118) on deduce the Maxwell's equations and the analogue of Maxwell's equations for gravitodynamics.

Resolving the equations system (118) by using a gauge transformation alike to Lorentz gauge, on obtain the wave equations of unified scalar and vector potential:

$$\Delta \Phi - \frac{1}{c^2} \frac{\partial^2 \Phi}{\partial t^2} = -\frac{r_\Phi}{e_\Phi}$$
$$\Delta \vec{A} - \frac{1}{c^2} \frac{\partial^2 \vec{A}}{\partial t^2} = -\frac{\vec{j}_\Phi}{e_\Phi}$$
(122)

or in quadridimensional notation:

$$\Delta A_m - \frac{1}{c^2} \frac{\partial^2 A_m}{\partial t^2} = -\frac{(j_\Phi)_m}{e_\Phi}$$
(123)

On see a perfect similarity between electrodynamics and the dynamics of unified field. On can extend this similarity to gravitodynamics.

**5.2. Lagrangean approach**

Following the model of demonstration from section 4, on write the Lagrangean for probe (mass $m_0$ and charge $e_m$) which moves in gravitational and electric fields, as follows:

$$L = -\frac{m_0 c^2}{k'} \cdot \sqrt{1 - \frac{v^2}{c^2}}$$
(124)

For k' given by (116) and weak field approximation ($\Phi \ll c^2$), results:

$$L \cong -m_0 c^2 \cdot \sqrt{1 - \frac{v^2}{c^2}} \cdot \left(1 + \frac{\Phi - \vec{v} \cdot \vec{A}_\Phi}{c^2 \cdot \sqrt{1 - v^2/c^2}}\right) = -m_0 c^2 \cdot \sqrt{1 - \frac{v^2}{c^2}} - m_0 \Phi + m_0 \vec{v} \cdot \vec{A}_\Phi$$
(125)

After some calculations result:

$$L = -m_0 c^2 \cdot \sqrt{1 - \frac{v^2}{c^2}} - m_0 \Phi_{grav} - e_m \Phi_{el} + m_0 \vec{v} \cdot \vec{A}_{grav} + e_m \vec{v} \cdot \vec{A}_{el}$$
(126)

Relationship (126) is exactly the Lagrangean of particle in electric and gravitational field. On emphasize that appear the term for gravitomagnetic field.

Using Lagrange equation and definitions from (117) on obtain the equation of probe motion in electric and gravitational fields:

$$\frac{d\vec{p}}{dt} = m_0 \vec{E}_u + m_0 \vec{v} \times \vec{B}_u$$
(127)

where the expression from right side is the unified field analogue of Lorentz force. From (127) on can take apart the Lorentz force and the gravitational analogue of Lorentz force.

**6. Additional effects**

In this section on see about the consequences of supplementary terms from (70) and (69). We'll discuss some implications of superior order approximations that were neglected in demonstration for unified potential. A more detailed discussion on this subject will be the subject of a future paper.

## 6.1. Consequences of additional terms from unified potential expression

On rewrite (70) for unified potential using relationships (51), (52) and (54) and the link between asymmetry and strength **E** of electric field (58):

$$\Phi(R) = -\frac{GM}{R} + \frac{e_m e_M}{4\pi\varepsilon_0 m_0}\frac{1}{R} + \frac{|e_m|r_0}{m_0}\overline{\cos\boldsymbol{\alpha}_B}E(R) - 2\pi\varepsilon_0\left|\frac{e_m}{e_M}\right|\frac{r_0}{m_0}R_0^2 E^2(R) \quad (128)$$

where M and $e_M$ characterize the source, $m_0$, $e_m$ and $r_0$ characterize probe and $E(R)=|\mathbf{E}(R)|$.
If $|e_M|=|e_m|=e$, then on write (128) as follows:

$$\Phi(R) = -\frac{GM}{R} \pm \frac{e^2}{4\pi\varepsilon_0 m_0}\frac{1}{R} + \frac{er_0}{m_0}\overline{\cos\boldsymbol{\alpha}_B}E(R) - 2\pi\varepsilon_0\frac{r_0}{m_0}R_0^2 E^2(R) \quad (129)$$

First two terms from right side of (129) are known: gravitational potential and electric potential multiplied with probe charge/mass ratio. If on apply the gradient with (-) sign, those two terms give the probe acceleration in gravitational and electrical fields of source.

Next, we focusing on the last two terms of right side of (129). For the beginning let's evaluate the constant factors of the two terms. If probe is the proton, then $m_0=1.673 \cdot 10^{-27}$ kg and $r_0=10^{-15}$ m. The average of $\cos\alpha_B$ between limits 0 and $\pi/2$, for hemispheric area, is 0.63. $R_0$ must be smaller than radius of the smaller steady particle (the electron). Because in TFM results that radius and mass of any particle is in proportion, it results that electron radius is 1836 times smaller than proton radius (this subject will be developed further). As result $R_0<4.5 \cdot 10^{-19}$ m. With these values results:

$$K_1 = \frac{er_0}{m_0}\overline{\cos\boldsymbol{\alpha}_B} \cong 4.8 \cdot 10^{-8}$$

$$K_2 = 2\pi\varepsilon_0\frac{r_0}{m_0}R_0^2 \leq 0.7 \cdot 10^{-35} \quad (130)$$

The last two terms of potential can be writing:

$$\Phi_1(R) = K_1 \cdot E(R)$$

$$\Phi_2(R) = -K_2 \cdot E^2(R) \quad (131)$$

And the corresponding probe accelerations are:

$$\vec{a}_1(R) = -K_1 \cdot \mathrm{grad}\, E(R)$$

$$\vec{a}_2(R) = K_2 \cdot \mathrm{grad}\left[E^2(R)\right] \quad (132)$$

Those supplemental accelerations determine supplemental forces. As result, Lorentz force on probe ($e_m$, $m_0$) must complete as follows:

$$\vec{F}_L = e_m\vec{E} + e_m\vec{v}\times\vec{B} - m_0 K_1 \mathrm{grad}\left|\vec{E}+\vec{v}\times\vec{B}\right| + m_0 K_2 \mathrm{grad}\left(\vec{E}+\vec{v}\times\vec{B}\right)^2 \quad (133)$$

If on append the forces due to gravitostatic and gravitomagnetic forces, on have the wholeness of forces that acting on probe.

A summary calculus of relative strength of the four acceleration components resulting from unified potential for the electron on first Bohr orbit in hydrogen show:

$$a_{grav} \cong -4.3 \cdot 10^{-17}\ ms^{-2}$$

$$a_{el} \cong -1 \cdot 10^{23}\ ms^{-2}$$

$$a_1 \cong 6.9 \cdot 10^{14}\ ms^{-2} \quad (134)$$

$$a_{2mxim} \cong -7 \cdot 10^{-4}\ ms^{-2}$$

On starting the discussion from $\Phi_2$ term of unified potential. On observe that from the three terms linked to interactions by asymmetry of fluctuations distribution, this one have the smallest contribution. It is always attractive and, at microscopic scale, it is important only at very small distances because its dependence of $R^{-5}$.

This term is due to gradient of asymmetry and that's why is always oriented from the zone with weak field to the zone with intense field. Both the charged and neutral particles are subject to the effect.

On not confuse this effect with the acceleration of a dielectric in electric field gradient, which is due to electrical interaction between dipole from polarized dielectric and electric field.

Let's take two systems: an asymmetric capacitor charged at a potential difference (or any other system that generate an electrostatic field gradient) and the second one that is a neutral body. According to above discussions, on neutral body act a force due to $\Phi_2$ that is oriented towards intense field, and no force act on field generator. It is exactly the situation described in section 3.5. that corresponds to Newton's third law violation. In the same time, the momentum conservation law is violated and in an adequate experimental configuration the angular momentum conservation law is violated too.

If the two systems are rigid linked, the whole ensemble accelerates on direction of field gradient increase. The acceleration continues as long as the gradient exists.

In the same time, although the internal energy is constant, the ensemble gains kinetic energy related to an external observer. It is possible that the gained energy overrun the energy needed for maintain the field gradient. In theory on consume energy only when capacitor is charged, but practically there are continuous leakage in dielectrics.

As consequence, the energy conservation law must reformulate as follows:

***For an isolated system, for the bodies between exert interactions of the same kind, the energy of system conserve.***

And for the angular momentum conservation law:

***For an isolated system, for the bodies between exert interactions of the same kind, the angular momentum of system conserve.***

For the violation of main conservation law are possible two explanations:

**A.** There is indeed situations in which conservation laws are not valid and it is necessary to complete them with their limits, as on reformulate the Newton third law and the conservation laws of momentum and energy. Up to now, the physics shows that the violation of a law implies a higher symmetry, together with an appropriate conservation law. On illustrate by parity violation that led to combine parity law conservation CPT;

**B.** It is possible that the two-body system (supposed to be isolated) it is not isolated in fact. In this situation, to preserve the validity of conservation laws, it must include the *hidden* part, as it was neutrino introduces for preservation of conservation laws in weak interactions.

What is the very nature of *hidden* part? Because the interactions between particles realize by fluctuations ensemble mediation and the particles are entities that not differ from fluctuations, it is genuine to suppose that the fluctuations ensemble could be (in certain conditions) part of a system such the one described above. In that situation the apparently violation of certain physical quantities is due to neglect the fluctuations ensemble as part of the system.

It is possible that electric field asymmetry produces by certain systems (such is the one described above) can realize an opening of those systems towards the fluctuations ensemble that represent the space-time continuum (or the vacuum). It is known that vacuum contain a huge amount of energy (so called zero point energy). If on include the vacuum as part of a system open the access to that energy. Shortly, the electric field asymmetry could bring the coupling of vacuum to system that generate the asymmetric electric field and what on see as conservation laws violation is due to ignore that vacuum is part of the system.

Another point of view: interaction between neutral body and the system that generate the asymmetric field is mediated by vacuum. The inequality of forces that the two systems act one each other could be produces by vacuum tension induced by field asymmetry.

From the point of view of thermodynamics, it is possible that the systems with symmetrical fields are closed systems at thermodynamic equilibrium, and the systems with asymmetrical fields are open systems, far from thermodynamic equilibrium and cross by energy flow. More, the newly observation of thermodynamics second law violation for mesoscopic systems [47] can suggest the following: the systems with asymmetry fields (higher degree of order) catalyze the local self-organization of vacuum fluctuations (non entropy process), that is equivalent with a energy transfer from vacuum to self-organized system and, consequently, to field asymmetry system. With reference to discussion from point A, those phenomenons can show that at the whole Universe scales the entropy on conserve.

Without neglect the possibility that the explanation A could be true, on consider that explanation B is more close to the correct one by means of bring arguments and because is hypothesis saving. The future will confirm one of the two hypotheses. However, apart from implications for knowledge, important is that, practically, the above discussion will bring applications in transport and energy production.

The practical use of $\Phi_2$ term effects is small because the constant factor $K_2$ is very small. The factor $grad(E^2)$ must be very large, at limits or beyond the present technological possibilities.

Let's focusing on the effects of $\Phi_1$ term from unified potential expression. This term act on electrical charged probe particles. If on analyze the behavior of this term in (50) or (69), on observe that irrespective of charge sign of probe or the orientation of **E**, the probe accelerate from the zone with intense field towards the zone with weak field.

Let's take the same system composed from asymmetric electric field generator and from a dielectric in that field. On every charge of the dielectric act a force with orientation from the zone with intense field towards the zone with weak field. The dielectric will not act at all on asymmetric electric field generator and so the whole system will self-accelerate. The discussion from $\Phi_2$ remain valid, the only distinction is the direction of movement.

If dielectric have the relative electric permittivity $\varepsilon_r$ then instead of constant factor $K_1$ on take $K_1\varepsilon_r$.

Because $K_1$ is much more large than $K_2$ it is possible that for comparative small values of gradient of electric field strength modulus (in the range of the actual possibilities of technology), on obtain the values for acceleration (and force) that are useful for transport and/or energy production.

The effect of $\Phi_1$ term is known as Biefild-Brown effect that is the emergence of a force with the same orientation as electric field gradient in an asymmetric capacitor with or without dielectric. There are patents for devices based on this effect ([3] – [6]). In the last years on offers patents ([7], [18], [37]) for similar devices, all with applications in transportation.

On the other hand, starting from observations on gases discharge (Chernetski) there are researches (finalize by patents [8] – [10], [38]) that dignify (in certain discharge conditions) the formation of abnormally accelerated charge clusters towards cathode. As result, in certain discharge conditions, the energy recovered from system was more than the energy needed for maintains the discharge.

There is the natural phenomenon of ball lightning: its stability and radiated energy have not found yet a convincing explanation.

The analysis of above phenomenon within TFM results can offer explanations using relationship (133) both for charge clusters and ball lightning cohesion (the coupling between vacuum and systems that generate asymmetrical electric field) and abnormal acceleration of charge clusters (probably in strong electric field gradient in the cathode vicinity).

In [14], [15] on show that the energy absorption process by the particle that is accelerated in electric field it is not independent of the particle movements: the energy absorption rate decrease with velocity increase and when the velocity approach c the absorption rate approach zero. This phenomenon was interpreted as inertia (and consequently masses) increase: for v→c results m→∞.

Remembering the above discussion on return to self-accelerated systems. Because asymmetric field source and dielectric are in relative rest and because the movement of system related to vacuum have not any meaning (otherwise the vacuum is an absolute reference frame), the apparently inertia increase is not produces anymore. Consequently, the system accelerates until field asymmetry cease. The light speed cease to be a barrier.

Concerning the SRT consequence that any body can't surpass the light speed, it must observe that the any mean for particle acceleration is based on electromagnetic interaction. As result of above discussion, the SRT consequence can be formulate as follows:

***Any body can't accelerate so that its relative speed related to accelerating field source surpasses the light speed.***

**6.2. Second order effects**

If in (40) on hold the second order terms from exponent development, the expression of acceleration and unified potential complicate. We not intent now to full develop this subject but we stop on a single consequence. For distances much more large than atomic dimensions, the contribution of superior terms with δ from opposite charges on cancel because matter is neutral at macroscopic scale. It remain only the terms from fluctuations concentration contribution, so the acceleration (69) and unified potential (70) will be greater with the factor:

$$1 + C\frac{R_s}{R} = 1 + \frac{GM}{c^2 R} \tag{135}$$

Consequences:

**a.** On gravitational interaction. The forces between two different bodies with masses $M_1$ and $M_2$, situated at distance R one each other are:

$$F_{1 \to 2} = \frac{GM_1 M_2}{R^2}\left(1 + \frac{GM_1}{c^2 R}\right)$$
$$F_{2 \to 1} = \frac{GM_1 M_2}{R^2}\left(1 + \frac{GM_2}{c^2 R}\right) \tag{136}$$

**b.** On electric interaction. The forces between two particles $(e_1, M_1)$ and $(e_2, M_2)$ situated at distance R one each other are:

$$F_{1 \to 2} = \frac{e_1 e_2}{4\pi\varepsilon_0 R^2}\left(1 + \frac{GM_1}{c^2 R}\right)$$
$$F_{2 \to 1} = \frac{e_1 e_2}{4\pi\varepsilon_0 R^2}\left(1 + \frac{GM_2}{c^2 R}\right) \tag{137}$$

On observe that the forces are not stringent equals, the body with largest mass act with a largest force. It is as the largest mass bodies have the largest electric charge. As for above, the discussion from section 6.1 concerning the conservation laws violation on applies for electric interactions too.

If electric charge depends on mass, then the proton charge is slightly large than the electron one, which signifies that the atom it is not quite neutral. Although the difference is small, at cosmological scale the effect could be significant. The newly findings that Universe expand more than expected can be explain by the slight excess of positive charge of any body that is due to proton-electron charges disparity.

On can conclude that conservation laws violation is a phenomenon encounter at all interactions, and that suggest (see the hypothesis from section 6.1) that any system is coupled with vacuum.

The superior order terms from exponent development in (39) become more important for subatomic distances. As consequence, the understanding of interactions at distance comparative with nucleus one need that those terms must enter in interaction's calculus. This subject is far beyond the subject of this paper. On conjecture that the two types of nuclear interactions can be explained using the superior terms from exponent development in (39). The nuclear forces act at small distances because the superior order terms are significant only for small distances. The success in explains nuclear forces by TFM will offer the key for the desired unitary field theory. For the reach of this goal, it must find an adequate shape of distribution function for fluctuations at small distances.

### 7. Open problems

In this section on broach the subjects that are insufficient developed because are beyond the purpose of this paper or/and are speculative. Some of this problem rise questions that worth time and effort investments because the answers could show another image of reality and allow for new technologies.

#### 7.1. A model for elementary particle

In the frame of TFM on loom a model for steady elementary particle, in that the particle is characterize by its radius that is in proportion with its mass. In the same time the fluctuations concentration on particle surface is the same whatever is the particle and the asymmetry of fluctuations velocity distribution is also maximum on its surface. The two possible orientations for velocity of fluctuations correlate to the two types of electric charges. Because the asymmetry of fluctuations velocity distribution is maximum on particle's surface whatever is the particle explain why the absolute value of charge is the same whatever is the particle.

The particle is steady perhaps due to the combination of the following factors: radius, the value of fluctuations concentration, the orientation of fluctuations velocity at surface and rotation around an own axis. It is possible that al those factors contribute to realize a certain value of gradient of average lifetime that this gradient allow to keep the cohesion of particle. The steady particle may be considered as a self-organization system that broad its influence outside by decreasing with distance modification of surrounding fluctuations field. The particle and the field are the same entity; the particle is linked with its field. By field mediation the particles influences one each other. The change in the movement status of a particle induces changes in the surrounding fluctuations field that propagate with finite velocity and are perceived as electromagnetic and/or gravitational waves. The relative motion of two particles make that the field of one particle perceived by the other one must be different from static case. This situation is described by unified quadripotential instead of unified scalar potential. The relative motion: straight, rotational or spin produces the corresponding modification of fluctuations field and this is described by magnetic and/or gravitomagnetic fields. The short-range effects of fluctuations field could explain the two kinds of nuclear forces.

Of course, the above description is far away from a coherent theory of elementary particles. It must explain firstly why, amongst fermions, only the proton and the electron (together with its antiparticles) are steady, why only for the two values of radius particles have lifetime practically infinite and for other radius values the particles aren't steady. It is credible the presumption that the self-organization is particularly dynamic than static process. On handle the particles as spherical oscillators and link this fact with wave behavior of particles. The particle radius belongs to a discrete spectrum in that only two solutions are steady and the rest are more or less unstable.

Another problem: the measurement of proton and electron radius performed by collisions on lead to closed values for its radius. In the TFM the ratio of the two radiuses is 1836. We answer by remembering that collisions parameters were calculated by Coulomb law extrapolate to distances comparable with atomic nucleus dimension. On seen that at that small distances the superior order terms from (39) have a great contribution, so the interaction changes. This could explain the shown difference.

In fact, the proton and electron radius could be smaller, but this can be verified only by a rigorous measurement of factor $K_1$.

**7.2. The neutron**

On know the mass of proton, electron and neutron. According to TFM there are following ratio between radios of the three particles:

$$\frac{r_p}{r_e} = 1836.12$$

$$\frac{r_n}{r_e} = 1838.65$$

(138)

Based on the above data, on can imagine the neutron as a proton with an electron that rotate around it at an apparently distance equals 2.5 electron radius from proton surface. Because the neutron radius is an average between proton radius and electron orbit radius, the last one must be greater than 1838.65 electron radius. On remember that at these distances the superior order terms from (39) have an important contribution.

The above model is consistent with the following facts:
- the magnetic moment of neutron corresponds to a negative particle spin although the neutron is without electric charge;
- the collision experiments on neutrons and the value of gyro magnetic factor $g_n$=-3.83 show a charge distribution inside neutron that consist in a positive core and a outward negative layer;
- there is no atomic nucleus without neutrons (except light hydrogen). This shows that neutrons play a decisive role in atomic nuclei cohesion. The above model for neutron allows understanding that cohesion is due to a sort of covalent *link* neutron-neutron and neutron-proton mediates by neutron's electron. This is a very simple picture because for the description of nucleus interactions on must use the superior terms of (39).

**7.3. Fields quantification**

Because temporal fluctuations are discrete entities and its distribution by orientation is discrete too, on conclude that fluctuations field is quantified. It results that changes in the state of field and particles are discrete. As result, it is genuine to suppose that all physical quantities are quantified. Because is the basic level of the world, the quantification of fluctuations field bring the electromagnetic and gravitational fields quantification.

**7.4. Links with quantum physics**

The Schrödinger undulatory mechanics, now a fundamentally side of quantum theory, did not allow the interpretation of wave associate to particle as physical wave because of obvious stability of particle and wave packet spreading. At the fundamentally level, the contradiction is due to difference between phase and group velocities of de Broglie wave. Because Schrödinger equation is based on group velocity of de Broglie wave, for keeping the Schrödinger equation it must reject the interpretation of de Broglie wave as physical wave. The problem was solved (Bohr) by interpretation of squared wave function as probability amplitude. This is a core of Copenhagen interpretation. This interpretation it is not without contradictions, if on allude to introduction of weird concept of wave function collapse at the measurement instant.

In [13] on show that when deduce the phase velocity of de Broglie wave was made the supposition that the whole particle energy is kinetic. This is due to the supposition that particle is a one-dimensional entity. At a closer look, the above lead to the next closed logical circle:

*If energy is kinetic energy, phase velocity of a de Broglie wave is not equal to mechanical velocity of a particle. If phase velocity does not equal mechanical velocity, a free particle cannot consist of a single wave of specific frequency. If it does not consist of a single wave, then the wave-features of a particle must be formalized as a Fourier integral over infinitely many partial waves. In this case, any partial wave cannot be interpreted as a physical wave. If the interpretation as a physical wave is not justified, then internal wave-features cannot be related to physical qualities. If they cannot be related to physical qualities, then internal processes must remain unconsidered. And if internal processes remain unconsidered, then energy of the particle is kinetic energy.*

If on consider that particle have internal (potential) energy too, and particle oscillate between maximum kinetic energy status and maximum potential energy status, then the whole particle energy is:

$$E = E_c + E_p = m|\vec{u}|^2 = h\nu \tag{139}$$

where **u** is particle velocity and ν the frequency of associated wave. On can write the phase velocity:

$$c_{phase} = \frac{\textbf{\textit{w}}}{|\vec{k}|} = \lambda \nu = \frac{h\nu}{|\vec{p}|} = \frac{m|\vec{u}|^2}{m|\vec{u}|} = |\vec{u}| \tag{140}$$

Therefore, the phase velocity of associated wave equals particle velocity and the wave packet is no need anymore. The particle-associated wave is monochromic. Considering the wave energy hν as total energy doesn't raise problems because, experimentally speaking, the energy can be measured by movement status or by its wavelength and the appropriate physical quantities (u, λ, ν) are linked.

As result, the particle oscillation with frequency ν (that depend on particle mass and velocity) between the two statuses is in fact the mass oscillation around an average value. Because in TFM mass is linked by particle radius, the mass oscillations signify the oscillations of particle radius around an average value.

Remember that in section 7.1 on suppose that the particle's stability is correlated with particle as spherical oscillator.

On speculate that if on succeed the synchronization of body's particles oscillation, then with a small force applied when the particles radius are minimum, the inertia is minimum and the body accelerate very strong.

**7.5. On stability of electron orbit around atomic nucleus**

The planetary model of Rutherford (verified by collision experiments between α particles and atoms) had called in question the validity of physics concepts: the electrodynamics state that any accelerated charge radiate energy and decelerate.

In these conditions the rotating electrons around nucleus and that suffer a centripetal acceleration must continuously radiate energy and finally fall on nucleus. However, atoms exist, are steady and don't radiate continuously energy. The dilemma was: Rutherford model is wrong (but the experiences shown the opposite) or electrodynamics laws (experimentally confirmed too) are wrong. By his postulates, Bohr decided for the second alternative that is accepted by contemporary physics: the electrodynamics laws loose its validity in micro world. This is the moment when quantum physics separate from classical physics. The first postulate of Bohr became a sort of sui-generis law of nature, otherwise the intraatomic movement of electrons become unexplainable, regardless on treat the movement classical or probabilistic.

In [29] on set the basis for total generalization of inertial reference frames in the framework of a new theory called gravitovortex. Its author demonstrates that: *The movement of rotation in a vortex conserve indefinitely in absence of any external forces. In that movement there are no centrifugal forces. As result, the vortex is a perfect inertial system.* There exists an infinity of inertial equivalent systems that moves in rotational uniform motion refer to vortex, called vortex-inertial systems by the author. The equation of movement for a particle (mass m and density ρ) along the radial direction in the vortex with surroundings density ρ' is:

$$m\frac{d^2R}{dt^2} = m\frac{C_a^2}{R^3} - m\frac{\boldsymbol{r}}{\boldsymbol{r'}}\frac{\Gamma^2}{4\boldsymbol{p}^2}\frac{1}{R^3} \qquad (141)$$

where $C_a$ is area constant and $\Gamma$ is the vortex intensity:

$$C_a = R\dot{\boldsymbol{q}} = Rv' \quad and \quad \Gamma = 2\boldsymbol{p}Rv \qquad (142)$$

v' is tangential part of particle's velocity and v is the same quantity for vortex.

First term of right side of equation (141) is centrifugal force and the second one is a force due to pressure gradient in vortex field (on notate $F_v$).

On write equation (141) as follows:

$$m\frac{d^2R}{dt^2} = m\frac{C_a^2}{R^3}\left(1 - \frac{\boldsymbol{r}}{\boldsymbol{r'}}\frac{\Gamma^2}{4\boldsymbol{p}^2 C_a^2}\right) = m\frac{C_a^2}{R^3}\boldsymbol{g} \qquad (143)$$

where γ parameter, that is constant for a given particle and movement, have a straight significance: differentiate the movement of particle from vortex movement.

If particle is part of vortex (ρ=ρ ) and have the same movement as the vortex ($\Gamma^2/4\pi^2=C_a^2$) it results γ=0. This value of γ characterizes the vortex itself. On observe that $md^2R/dt^2=0$, so the movement is inertial. The same result is possible for a particle that is not part of vortex, but the ratio between vortex intensity and area constant have a value that determine γ to be zero.

The author demonstrate that for a particle which moves in a central field, where source of field generate also a gravitovortex field, the centripetal force is the same as $F_v$ and it is not correct to identify the centripetal force by central force. The central force (gravitational or electrostatic) characterizes the static interaction between the two particles meanwhile $F_v$ is due to slightly inequality between proton and electron charges. As result, if the atomic nucleus generate gravitovortex field too, it is possible that electron which moves on orbit around nucleus fulfill the conditions for γ=0. Results that electron movement is inertial and therefore it will not radiate energy continuously.

On see in section 6.2 that in TFM on obtain exactly such a mass dependence on charge and the supplemental force is in proportion with $R^{-3}$ as for $F_v$.

At atomic scale, if apart from electrostatic interaction on take the interaction bring by $\Phi_1$, it results a supplemental force that is in proportion with $R^{-3}$. Remember that $\Phi_1$ generate always a repulsive force that is in proportion with probe mass, the resulting effect of the two interactions is that repulsion between two protons is greater than the repulsion between two electrons. In the same time, in the case of proton-electron attraction, from electrostatic force that acts on proton must subtract a greater term than for the electron. The effect is much more stronger than the one described at section 6.2, and therefore is much more credible that supplemental force $F_v$ is given by this one.

**7.6. Beyond Special Relativity**

The SRT second postulate which states that light speed is the same in any inertial reference frame (that is independent of its source velocity) generate much discussions and was subject of many experimental validations. As C. Nordmann (quoted in [2]) wrote: *However, it remains something exciting in Einstein's system. It is very coherent but also supports on a peculiar conception of light propagation. How on imagine that propagation of the same light ray it's identical for an observer that run from it, as for an observer that come towards it? If this is possible, this is incredible by ours ancestral mentality, and we can't represent the nature of this propagation by any effort we do. Let's confess: here it is a mystery. The whole Einstein's synthesis, however coherent is, it support on a mystery, as in inspired religions.*

Because the idea of light speed independence of the movement of its source was difficult to assimilate, on made frequent trials for observed phenomenon explanation, without the second postulate. The most remarkable was the Ritz electrodynamics where the two homogeneous Maxwell's equations were kept, but the two others (that contain sources) was modified so the light speed is c only when is measured related to source. The theory is in correspondence with observations on stars position aberration, Fizeau experiment and Michelson-Morley experiment. Nevertheless, on quote The Michelson-Morley experiment with extraterrestrial light sources (stars, Sun) as experimental proof of the second postulate.

The second postulate offers to the light an unusual characteristic. The justification for second postulate on quote the negative result of the Michelson-Morley experiment, that can be also explained in agreement with mechanics laws by ballistic hypothesis (the Ritz theory). On show that light seem to propagate in all directions with the same velocity c due to supplemental effects that compensate the drag instead of light speed independence from source velocity. On the other hand, the contraction hypothesis of Lorentz (used for explain the negative result of the Michelson-Morley experiment) must be supplementary verified by optical and electrical experiences. Thus, Lord Rayleigh (1902) and Brace (1904-1905) researched the optical anisotropy that Lorentz contraction must produce; the result was completely negative. Because the wire electrical resistance depends on its length, Trouton and Rankine (1908) measured with a very sensitive Wheatstone bridge the electrical resistance of metallic wires that was parallel and then perpendicular on Earth movement. The result was negative too.

It is believed that relativity principle (first postulate) is violated if the light speed is not the same in all inertial reference frames. In fact, on confuse two different idea: the light speed c and the law x=ct for propagation of light. When on pass from an inertial reference frame (x=ct) to another, on can write x'=ct' or x'=c't. Both of relationships satisfy the relativity principle because each of them conserves the shape of the law for first reference frame. The relativity principle doesn't impose the constancy of light speed; it request that the law of light propagation must preserve its shape when on pass from an inertial reference frame to another one. As well, for eliminate the twin paradox it must that duration of the same event must be the same when measured in different inertial reference frames. It was tried to solve the paradox by considering that for the duration compare one reference frame must become noninertial (at least for a while) doesn't solve the problem because on exit from the frame of SRT. In [2] on demonstrate that duration dilatation from SRT is a direct consequence of inertial reference frame synchronization. It isn't a spontaneous dilatation made from nature but one that observers produce by clocks synchronization.

On the other hand, the extinction theorem (Ewald and Oseen) states that if an incidence electromagnetic wave that propagate with c velocity enters in a disperse medium, its field is cancelled by a part of induced dipoles and is replaced by another wave that propagate with medium corresponding phase velocity. Thus, the incident wave is cancelled by interference and replaced with another wave. For wave groups that reach over limit surface and begin propagate in medium, it is credible that the wave extinction and its replacement with the wave that correspond to the medium take place on a finite distance. For visible light, this distance is around $2*10^{-6}$ cm in glass, 0.04 cm in air and 2 light years in the interstellar medium. On see that whatever is the light speed then it is emitted, a interposed medium between source and observer replace the emitted radiation by a perturbation with same frequency, but with phase velocity that characterize the rest medium. Then, an observer in rest will measure a light speed equals c, the source movements and light speed relative to it are without signification. In this way, all older proofs and some experiences concerning the second postulate are vitiate by extinction theorem.

J. D. Jackson [17] considers that conclusive experiment, which confirmed the second postulate and that surpass the extinction theorem, was performed in 1964 at CERN, Geneva. The speed of 6 GeV photons produced by high-energy neutral pions disintegration was measured using time of flight on 80 m distance. Pion was produced by bombardment of beryllium target with 19.2 GeV protons with speed 0.99975*c. In the limit of experimental errors, the photons speed emitted by source in movement was measured and result is c. The correct value of extinction distance for 6 GeV photons is questionable, but is far greater than 100 m. On that basis, it was considered that the second postulate of SRT was verified.

On can formulate the following objection: the ordinary electromagnetic radiation sources are accelerating charges and the source from CERN experiment is the disintegration of neutral pions. If on refer to TFM (although we don't discuss the particles structure) it is possible that the fluctuations field perturbation produces by accelerate charges can be different from perturbation due to $\pi^0$ disintegration, although both of them are received as electromagnetic waves.

In TFM on consider true both SRT postulates but the model can be modified so the second postulate isn't necessary. In absence of fluctuations distribution data, the new model (only in qualitative blueprint shape) is highly speculative.

On introduce following changes:

1. in every infinitesimal vacuum region the fluctuations have velocities between 0 and a maximum value (much more greater than c and that can be taken ∞). The distribution by velocity modulus is homogeneous. An observer from any reference frame perceive practically the same fluctuations distribution, thus on obtain the condition that there isn't exist an absolute reference frame. It must emphasize that the value of limit velocity can't be ∞ and because of this reason the observers from different reference frames will perceive slight differences in fluctuations distribution. If the reference frame velocities are much more small than the limit velocity (and at astronomical scale wasn't observed very large velocities) the differences can be under the measurement precision;

2. the maximum interaction between fluctuations produce when their relative velocity is 0 and decrease as the relative velocity increase. For this reason the shape of quantity a from interaction potential calculation (relationship (18) and the followings) can be:

$$a = a_0 e^{-u \cdot v_{rel}^2} \quad sau \quad a = a_0 e^{-u \cdot |v_{rel}|} \tag{144}$$

where $a_0$ and u are constant and $\mathbf{v}_{rel}$ is the relative velocity of interacting fluctuations. If u has a great value, the fluctuations will interact strongly for $\mathbf{v}_{rel}$ closed to 0 and for $\mathbf{v}_{rel}$ great values the interaction can be neglected;

3.     at particle's surface the fluctuations have average velocity c related to reference frame with particle in rest. On expect that distribution function of fluctuations upon velocity modulus must have the shape from figure 7. As the distance from particle increase, the graphic of distribution function become more flat so at infinity it is independent of velocity.

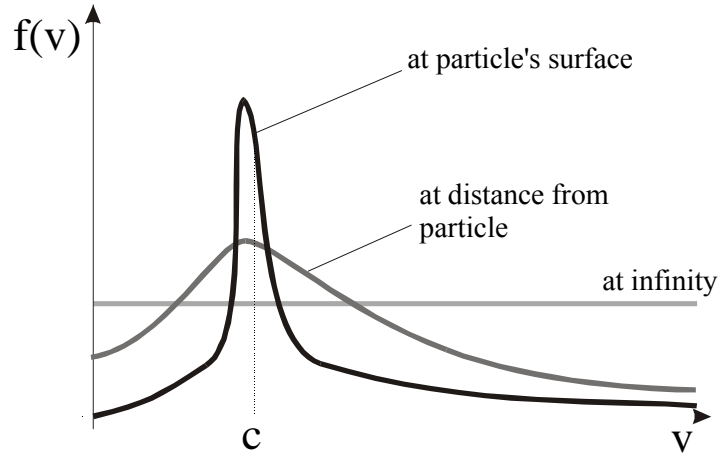

Figure 7

On take two charged particles (source S and probe P) that are in relative rest. The interaction between the two particles produces in the same manner as was described in section 3 with observation that the maximum of fluctuations distribution coincide, both for asymmetry δ produces by S and for Δ produces by P (in every point on the surface of P). In this situation, the interaction between S and P is maximum. If the two particles moves one each other with relative velocity $v_{rel}$ the maximum of the two distributions suffer a postponement and the strength of interaction decrease (figure 8). As the relative velocity increases, the interaction strength decreases.

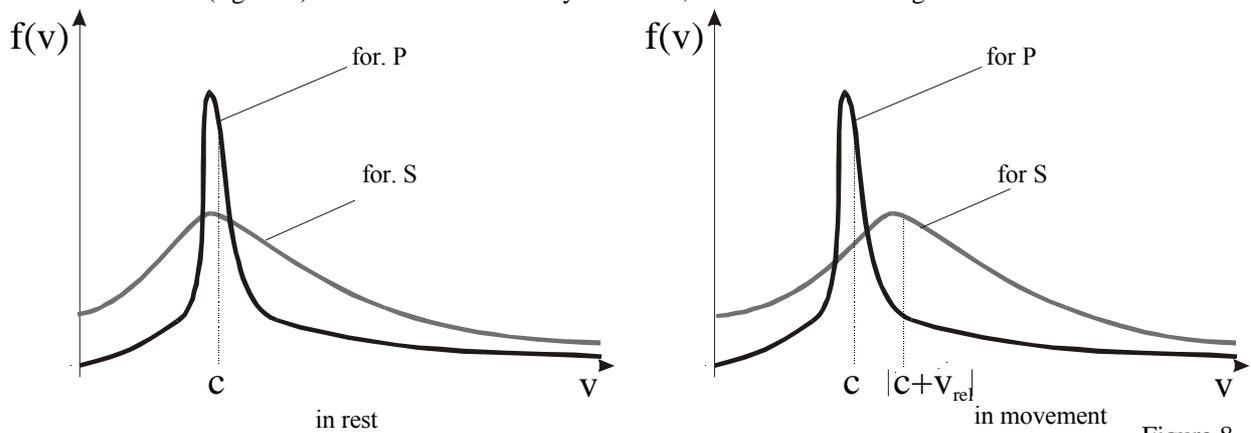

Figure 8

With a proper relationship for quantity a from (144) and with proper shape of distribution function, on could obtain for the factor that decrease the value of interaction potential (and P acceleration too) an expression closed to $(1-v_{rel}^2/c^2)^{1/2}$. The decrease of interaction between charges with increase of velocity was misunderstanding as mass (and consequently inertia) increase. The above new interpretation is in correspondence with conclusion (from section 6.1) that the velocity can exceed light speed but the relative velocity to the accelerator system can't exceed light speed. The same link between interaction strength and relative velocity can be valid for gravitational interactions.

On emphasize that above certain value of relative velocity (perhaps c) the bodies practically do not interact.

On see some consequences of the above model:

1. the model allow the velocities greater than light speed for bodies with mass greater than zero;

2. the bodies with relative velocity greater than light speed do not interact and, practically, do not exist one for the other. It is possible that two of these bodies pass one **through** other without visible reciprocal influence. Advancing with speculations, on imagine that an observer with superluminal velocity related to our galaxy (for instance) can't observe the familiar cosmic bodies, but he can observe another bodies that his velocity is smaller than c. It is as he gets into another universe. If he continues to accelerate, the *new universe* disappeared and the other will become visible because the observer has subluminal velocity, and so on. The successive *universes* can be completely distinct (meaning that there are domain for velocity values where the observer don't meet bodies), or can penetrate. But this is already a cosmological problem that surpass the intention of present paper;

3. on can suppose that fluctuations average velocities at particle's surface can be one of the governing facts for the particle stability (see discussion from section 7.1). It is possible that another combination of values (for particle radius, concentration and distribution of fluctuations at particle's surface, the stability produces by rotation around own axis and the average value of fluctuations velocity at particle's surface) determine another set of elementary particles that have another average fluctuations velocity at particle's surface, let's say $c_1$. This new set with interactions and physical laws alike to ours set frame too a different *universe*. It is natural to suppose that $c_1 \gg c$ so the interaction between new set of particles and *our* set will be minimum or even non-existent. As result, our universe and the new one can coexist **in the same space**, even if the relative velocities of bodies from the two universes are subluminal. On speculate further by supposing that are possible many particle sets, which correspond to another values of average fluctuations velocities $c_2$, $c_3$, .... $c_n$, and all corresponding to *universes* that coexist in the same space. A discrete spectrum of velocities $c_2$, $c_3$, .... $c_n$ indicate a quantification of vacuum potential, our *universe* being on the one of the levels of existence of vacuum. If exists some interactions between those universes, on should observe that our one seem to have more matter than the astronomical observations shows. However, this fact was already observed and determine the astrophysicists to deal with *dark matter* on its very nature on speculate.

**Bibliography**

[1] H. Aspden, "Ion accelerators and energy transfer processes", Patent GB 2002953

[2] N. Barbulescu, "Physical background of Einstein's relativity", Scientific Publishing House, Bucharest, 1979 – in romanian

[3] T.T. Brown, "Electrokinetic apparatus", Patent US 2949550

[4] T.T. Brown, "Electrokinetic transducer", Patent US 3018394

[5] T.T. Brown, "Electrokinetic generator", Patent US 3022430

[6] T.T. Brown, "Electrokinetic apparatus", Patent US 3187206

[7] J. Campbell, "Apparatus and method for generating thrust using a two dimensional, asymmetrical capacitor module", Patent US 6317310

[8] P. Correa, "Electromechanical transduction of plasma pulses", Patent US 5416391

[9] P. Correa, "Energy conversion system", Patent US 5449989

[10] P. Correa, "Direct curent energized pulse generator utilizing autogenous cyclical pulsed abnormal glow discharges", Patent US 5502354

[11] V. Drafta, "Nature is simple in essence – blueprint for unification of gravity and electromagnetism", http://xxx.lanl.gov/physics/0105095

[12] R.P. Feynman, R.B. Leighton, M. Sands, "Lectures in physics" (3 Volumes), Technical Publishing House, Bucharest, 1969-1971 – in romanian

[13] W.A. Hofer, "Beyond uncertainty: the internal structure of electrons and photons", http://xxx.lanl.gov/quant-ph/9611009

[14] W.A. Hofer, "Electron acceleration due to photon absorbtion: a possible origin of the infinity problems in relativistic quantum fields", http://xxx.lanl.gov/quant-ph/9805061

[15] W.A. Hofer, "A realist view of the electron: recent advances and unsolved problems", http://xxx.lanl.gov/quant-ph/9910036

[16] V. Hushwater, "Does the charge of a body reduce its gravitational field?" http://xxx.lanl.gov/gr-qc/0103001

[17] J.D. Jackson, "Classical Electrodynamics", (2 Volumes), Technical Publishing House, Bucharest, 1991 – in romanian

[18] A. Lafforgue, "Systemes isole auto-propulse par des forces electrostatiques", Patent FR 2651388

[19] L. Landau, E. Lifchitz, "Theorie des champs", Edition Mir, Moscou, 1970

[20] T. Mahood, "A Torsion Pendulum Investigation of Transient Machian Effects", http://www.serve.com/mahood/thesis.pdf

[21] C.J. de Matos, M. Tajmar, "Advance of Mercury Perihelion Explained by Cogravity", http://xxx.lanl.gov/gr-qc/0005040

[22] C. Moller, "The theory of relativity", University Press, Oxford, 1962

[23] L.B. Okun, K.G. Selivanov, "On the interpretation of the redshift in a static gravitational field", http://xxx.lanl.gov/physics/9907017

[24] L.B. Okun, "Photons and static gravity", http://xxx.lanl.gov/hep-ph/0010120

[25] L.B. Okun, "A thought experiment with clocks in static gravity", http://xxx.lanl.gov/hep-ph/0010256

[26] L.B. Okun, "Photons, clocks, gravity and the concept of mass", http://xxx.lanl.gov/physics/0111134

[27] Oleinik V.P., Borimsky Yu.C., Arepjev Yu.D., "Time, what is it? Dynamical Properties of Time", http://xxx.lanl.gov/quant-ph/0010027

[28] V. Ougarov, "Theorie de la relativite restreinte", Edition Mir, Moscou, 1979

[29] I.N. Popescu, "Gravitation", Scientific Publishing House, Bucharest, 1982 – in romanian

[30] E. Purcell, "Electricity and magnetism – The Berkeley lectures on physics", Didactics Publishing House, Bucharest, 1982 – in romanian

[31] Vesselin Petkov, "On the possibility of a propulsion drive creation through a local manipulation of spacetime geometry", http://xxx.lanl.gov/physics/9805028

[32] Vesselin Petkov, "Acceleration-dependent electromagnetic self-interaction effects as a basis for inertia and gravitation", http://xxx.lanl.gov/physics/9909019

[33] Vesselin Petkov, "Anisotropic velocity of light in non-inertial reference frames", http://xxx.lanl.gov/gr-qc/9909081

[34] Vesselin Petkov, "Is the active gravitational mass of a charged body distance-dependent?", http://xxx.lanl.gov/gr-qc/0104037

[35] H.E. Puthoff, "Polarizable-Vacuum (PV) representation of general relativity", http://xxx.lanl.gov/gr-qc/9909037

[36] F. Rohrlich "Classical charged particles", Addison-Wesley Publishing Comp., 1965

[37] H. Serrano, "Propulsion device and method employing electric fields for producing thrust", Patent WO 00/58623

[38] K. Shoulders, "Energy conversion using high charge density", Patent US 5018180

[39] M. Tajmar, C.J. de Matos, "Coupling of Electromagnetism and Gravitation in the Weak Field Approximation", http://xxx.lanl.gov/gr-qc/0003011

[40] A. Vankov, "Does light gravitate?", http://xxx.lanl.gov/astro-ph/0012050

[41] A. Vankov, "Proposal of experimental test of GRT", http://xxx.lanl.gov/gr-qc/0105057

[42] W.G. Unruh, "Time, Gravity and Quantum Mechanics", http://xxx.lanl.gov/gr-qc/9312027

[43] K. van Vlaenderen, "A charged space as the origin of sources, fields and potentials", http://xxx.lanl.gov/physics/9910022

[44] J.F. Woodward, T. Mahood, "Mach's Principle, Mass Fluctuations and Rapid Space-time Transport", http://chaos.fullerton.edu/Woodward.html

[45] R. Ziolkowski, M. Tippett, "Collective effect in an electron plasma system catalized by a localizad electromagnetic wave", Phys. Rew. A, 43-6, 1991

[46] R. Ziolkowski, "Electromagnetic localized wave that counteract Coulomb repulsion to catalize a collective electron-packet state", Phys. Rew. E, 52-5, 1995

[47] * * *, "Second Law of Thermodynamics Violated", Scientific American News, 24 july 2002